\documentclass[10pt,twocolumn,journal]{IEEEtran}

\usepackage{amsmath}
\usepackage{amssymb}
\usepackage{amsfonts}
\usepackage{amsthm}
\usepackage{enumitem}
\usepackage[ruled]{algorithm2e}
\usepackage{algpseudocode}
\usepackage{comment}
\usepackage{epsfig}
\usepackage{graphicx}
\usepackage{color}
\usepackage{cite}
\usepackage{nomencl}
\usepackage{xstring}
\usepackage{xstring}
\usepackage{xpatch}
\usepackage[utf8]{inputenc} 
\usepackage{graphicx} 
\usepackage{cleveref}
\usepackage{multirow}
\usepackage{mathrsfs}  

\theoremstyle{plain}
\newtheorem{thm}{Theorem}
\newtheorem{lem}{Lemma}

\theoremstyle{definition}

\title{FDD Massive MIMO: How to Optimally Combine UL Pilot and Limited DL CSI Feedback?
\thanks{Jungyeon Kim is with Department of Electrical Engineering, POSTECH, Pohang, South Korea (e-mail: jungyeon.kim@postech.ac.kr).}
\thanks{Jinseok Choi is with School of Electrical Engineering, KAIST, Daejeon, South Korea (e-mail:jinseok@kaist.ac.kr).}
\thanks{Jeonghun Park is with School of Electrical and Electronic Engineering, Yonsei University, Seoul, South Korea (e-mail: jhpark@yonsei.ac.kr).}
\thanks{Ahmed Alkhateeb is with School of Electrical, Computer, and Energy Engineering, Arizona State University, Tempe, AZ 85281 USA (e-mail: alkhateeb@asu.edu).}
\thanks{Namyoon Lee is with School of Electrical Engineering, Korea University, Seoul, South Korea (e-mail: namyoon@korea.ac.kr).}
}

\author{\IEEEauthorblockN{Jungyeon~Kim},
 {\it Graduate~Student~Member,~IEEE},
\and
\IEEEauthorblockN{Jinseok~Choi},
 {\it Member,~IEEE},
\and 
\IEEEauthorblockN{Jeonghun~Park},
 {\it Member,~IEEE},
\and 
\IEEEauthorblockN{Ahmed~Alkhateeb},
 {\it Senior~Member,~IEEE},
\and 
\IEEEauthorblockN{Namyoon~Lee},
{\it Senior~Member,~IEEE}
}

\begin{document}

\maketitle
 
\begin{abstract}
In frequency-division duplexing (FDD) multiple-input multiple-output (MIMO) systems, obtaining accurate downlink channel state information (CSI) for precoding is vastly challenging due to the tremendous feedback overhead with the growing number of antennas. Utilizing uplink pilots for downlink CSI estimation is a promising approach that can eliminate CSI feedback. However, the downlink CSI estimation accuracy diminishes significantly as the number of channel paths increases, resulting in reduced spectral efficiency. In this paper, we demonstrate that achieving downlink spectral efficiency comparable to perfect CSI is feasible by combining uplink CSI with limited downlink CSI feedback information. Our proposed downlink CSI feedback strategy transmits quantized phase information of downlink channel paths, deviating from conventional limited methods. We put forth a mean square error (MSE)-optimal downlink channel reconstruction method by jointly exploiting the uplink CSI and the limited downlink CSI. Armed with the MSE-optimal estimator, we derive the MSE as a function of the number of feedback bits for phase quantization. Subsequently, we present an optimal feedback bit allocation method for minimizing the MSE in the reconstructed channel through phase quantization. Utilizing a robust downlink precoding technique, we establish that the proposed downlink channel reconstruction method is sufficient for attaining a sum-spectral efficiency comparable to perfect CSI.

\end{abstract}

\section{Introduction}\label{Sec: Introduction}
\subsection{Motivation}
Massive multiple-input multiple-output (MIMO) is recognized as a pivotal technology for the future of wireless communications, with its focus on achieving high spectral efficiency. It has been primarily explored within the context of time-division duplexing (TDD) massive MIMO systems \cite{marzetta2010noncooperative, marzetta2015massive, ngo2013energy}, where its cornerstone feature is channel reciprocity. In TDD systems, both uplink (UL) and downlink (DL) channels are reciprocal, enabling the acquisition of channel state information at the transmitter (CSIT) without the need for additional DL channel training or feedback. This reciprocity simplifies system design and enhances energy and spectral efficiency.  

Unfortunately, many real-world wireless systems operate on frequency-division duplexing (FDD) rather than TDD. The preference for FDD is not solely due to regulatory standards; it offers technical benefits such as lower latency and more stable BS transmit power, which enhances amplifier linearity. In FDD massive MIMO systems, however, accurate CSIT acquisition at a base station (BS) requires an extremely high demand for training and feedback overhead due to the lack of channel reciprocity \cite{jindal2006mimo, lee2013space, bjornson2016massive, love2005limited, rao2014distributed,NLee_limited}. These overheads could potentially diminish the benefits of massive MIMO systems in FDD setups \cite{bjornson2016massive}.

 
Recently, significant progress has been made in reducing channel state information (CSI) feedback overhead in FDD systems, which include compressive sensing (CS)-based methods \cite{rao2014distributed, gao2015structured, han2017compressed, liang2020deep}, hybrid precoding methods that harness channel spatial correlation \cite{alkhateeb2014channel, tian2020randomized}, and DL channel extrapolation using UL pilots \cite{alrabeiah2020deep}. Notably, channel extrapolation to estimate the DL channel with UL pilots is gaining traction, presenting an appealing solution to eliminate the substantial signaling overhead required for DL channel training and CSI feedback. Consequently, there has been a surge in research exploring the integration of channel extrapolation with deep learning \cite{yang2020deep, wang2018deep, sohrabi2021deep}. However, accurately estimating the DL channel using UL pilots presents a significant challenge, notably due to the absence of channel reciprocity.

Despite the absence of channel reciprocity, certain geometric parameters of the channel, such as the angle of arrival (AoA), angle of departure (AoD), and channel path gain, can still be effectively utilized due to their (quasi) frequency-invariant properties. This fact allows for partial channel reciprocity to be leveraged in reconstructing DL channels  \cite{vasisht2016eliminating, dai2018fdd, ding2018dictionary, khalilsarai2018fdd}.  However, the accuracy of such DL channel reconstructions diminishes when there is a significant frequency discrepancy between UL and DL channels or when the number of channel multipaths increases. This necessitates adjustments to compensate for the discrepancy \cite{han2023fdd}. In this paper, we explore a critical question: How can the BS effectively estimate the DL channel by integrating UL pilot signals with DL CSI feedback? Our major finding indicates that transmitting a few bits of information, which quantize the phase of the channel paths to minimize the mean square error (MSE) in DL channel reconstruction, significantly enhances the DL sum spectral efficiency, especially when combined with the frequency-invariant channel parameters derived from the UL pilots.




\subsection{Related Works}

The DL channel estimation problem from UL pilots, which is also known as the channel extrapolation problem, has been extensively studied with various methods to diminish the DL CSI feedback overhead in FDD massive MIMO  \cite{vasisht2016eliminating, khalilsarai2018fdd, han2023fdd, yin2022partial, han2019tracking}. In \cite{khalilsarai2018fdd}. The underlying approach to solving the channel extrapolation problem is to leverage the shared channel structure in UL and DL. To be specific, UL and DL channels are typically modeled by the weighted sum of array response vectors created by AoAs and AoDs. The weight parameters are called complex-valued path gains comprised of the path amplitude and phase. The UL and DL phases per path are functions of their carrier frequencies; they can be modeled with statistical correlation \cite{vasisht2016eliminating,han2023fdd} or independence \cite{yin2022partial,adhikary2013joint,khalilsarai2018fdd,han2019tracking,qin2022partial,dai2018fdd,ding2018dictionary}.  
  
Leveraging this shared geometric structure in UL and DL channels, several channel extrapolation algorithm was proposed in \cite{khalilsarai2018fdd,dai2018fdd,yin2022partial,adhikary2013joint,vasisht2016eliminating,han2023fdd}. Using the fact that angular scattering function of the channels is invariant over wavelengths, the DL channel covariance estimation method was investigated. For instance, the  DL channel covariance matrix estimation technique was proposed using UL pilots in \cite{khalilsarai2018fdd}.  In \cite{yin2022partial}, a codebook based on partial channel reciprocity has been proposed and showed that the number of feedback bits can be reduced under a low-rank condition of the channel covariance matrix. 

Instead of estimating the DL channel covariance matrix, a direct DL channel reconstruction algorithm using UL pilots was introduced \cite{vasisht2016eliminating, rottenberg2020performance, han2023fdd}. This DL channel reconstruction approach leverages frequency-invariant parameters, such as AoAs and path gains, shared between the UL and DL channels. Specifically, in \cite{rottenberg2020performance}, a high-resolution channel reconstruction method was proposed, deriving the Cramér-Rao lower bound of the mean squared error (MSE) to delineate the limit of reconstruction performance. Additionally, previous studies have explored integrating this method with techniques for predicting channel parameters in high-mobility environments to enhance DL channel reconstruction \cite{qin2022partial, han2019tracking}. Very recently, an MSE-optimal DL channel reconstruction method was introduced in \cite{han2023fdd}, utilizing UL pilots. The key idea of this MSE-optimal method is to reconstruct the DL channels by harnessing AoAs and path gains derived from UL pilots, along with the phases for each channel path, aiming to minimize the MSE, assuming that the phases of UL and DL channels are statistically correlated, as suggested in \cite{vasisht2016eliminating}. The MSE analysis revealed that the accuracy of DL channel reconstruction diminishes with an increase in the number of channel paths or a significant discrepancy between UL and DL carrier frequencies. 
 
With the advancement of deep learning techniques capable of estimating nonlinear functions, there has been a surge in research integrating deep learning with channel extrapolation \cite{wang2018deep, alrabeiah2020deep, alrabeiah2019deep, yang2020deep, zhang2020cv, alkhateeb2019deepmimo}. In \cite{alrabeiah2019deep}, a mapping function that takes UL channels as input and produces DL channels as output was learned based on deep learning, under the condition that this function is bijective. In \cite{alrabeiah2020deep, alkhateeb2019deepmimo}, a method that leverages deep learning to predict DL codebooks using UL CSI has been proposed, utilizing the DeepMIMO dataset which is a pre-generated ray-tracing channel dataset. Additionally, from the perspective of using supervised learning, a method based on generative adversarial networks (GAN) has been proposed to predict the DL channel or its covariance \cite{ banerjee2022downlink}.  To apply the GAN image processing methods, the channel or its covariance matrix is converted into a specialized image format used for training the network. These deep learning-aided UL-DL channel mapping methods have presented new possibilities to enable massive MIMO FDD systems. However, these methods are highly sensitive to changes in the channel distribution, necessitating that the channel distribution during the test period remains consistent with that during the training period. This fact makes challenge to apply in practical FDD massive MIMO systems yet. 

Relying solely on UL pilots for DL channel reconstruction is limited when the phases of UL and DL channel paths are statistically independent as shown in the MSE analysis \cite{han2023fdd}. To address this issue, the DL channel reconstruction algorithm presented in \cite{Ugurlu2016} utilizes both UL pilots and multipath phase feedback from the DL channel. This method is attractive because the BS can enhance the accuracy of DL channel reconstruction by combining the DL multipath phase information along with the AoAs and path gains extracted from the UL pilot. However, when the amount of CSI feedback is fixed, the literature in \cite{Ugurlu2016} has not explored the optimal feedback bit allocation strategy for the multipath DL phase and the robust multi-user MIMO (MU-MIMO) precoding method resilient to the CSI feedback error. 


\subsection{Contributions}
We summarize the contributions of this paper as follows:
\begin{itemize}

\item We introduce the optimal CSI feedback bit allocation strategy for multipath phase information when the total amount of DL feedback information is fixed. Assuming the BS can estimate frequency-invariant parameters such as AoAs and path gains, the DL user sends back the phases of the multipath channel components using a scalar quantizer via a limited feedback channel. The proposed CSI feedback bit allocation strategy is to allocate variable bits per channel path according to the channel path gain to minimize the MSE. To accomplish this, we characterize the MSE between the true DL channel and the reconstructed DL channel with the quantized multiphase information as a function of quantization bits allocated to each channel path. Then, we present the MSE-optimal CSI feedback bit allocation strategy under a fixed total number of CSI feedback bits. The primary finding is that allocating more feedback bits to channel paths with higher path gain minimizes the MSE.

\item  Utilizing the developed MSE-optimal CSI feedback strategy for multipath phase information, we introduce an MSE-optimal DL reconstruction algorithm that combines the limited DL phase information with the AoAs and the path gains of the DL, which are extracted from UL pilots. Our reconstruction algorithm extends the previous work by optimally combining UL pilot information with DL CSI feedback to minimize the MSE of DL channel reconstruction and operates for the more generalized channel assumption than that in \cite{han2023fdd}.

\item To provide a comprehensive framework from channel estimation to precoding, we also introduce a robust DL precoding based on a generalized power iteration method \cite{choi2019joint} by utilizing the reconstructed channel and its error covariance matrix derived in the error analysis. The proposed algorithm leverages the MSE matrix from DL channel reconstruction to perform robust precoding against reconstruction errors. Moreover, it guarantees a locally optimal solution for joint user selection, power allocation, and beamforming to maximize DL sum spectral efficiency.

\item From simulations, we demonstrate that the proposed robust precoding method combined with our DL channel reconstruction technique, even with limited CSI feedback bits per channel path, achieves sum-spectral efficiency comparable to the weighted minimum MSE (WMMSE) precoding technique that uses perfect CSIT \cite{christensen2008weighted}. These results indicate that near-optimal DL sum spectral efficiency is achievable with a small amount of DL CSI feedback for the multipath phase, especially when it is optimally combined with UL pilot information. The required CSI feedback bits do not scale with the number of BS antennas but rather with the number of \textit{dominant channel paths.}
    
\end{itemize}

\section{System Model}\label{Sec: System Model}
In this section, we explain an FDD massive MIMO system model where a BS is equipped with $N$ antennas and $K$ users each have a single antenna. Our focus is on a single-cell network to deliver a clear explanation of the proposed method as in most prior work \cite{khalilsarai2018fdd, dai2018fdd, yin2022partial, adhikary2013joint,  han2023fdd, vasisht2016eliminating, rottenberg2020performance, wang2018deep, alrabeiah2020deep, alrabeiah2019deep, yang2020deep, zhang2020cv, alkhateeb2019deepmimo}. However, the proposed method can be adapted to multi-cell networks with minor modifications to account for pilot contamination effects. 

\begin{figure}[t]
	\centering
    \includegraphics[width=\linewidth]{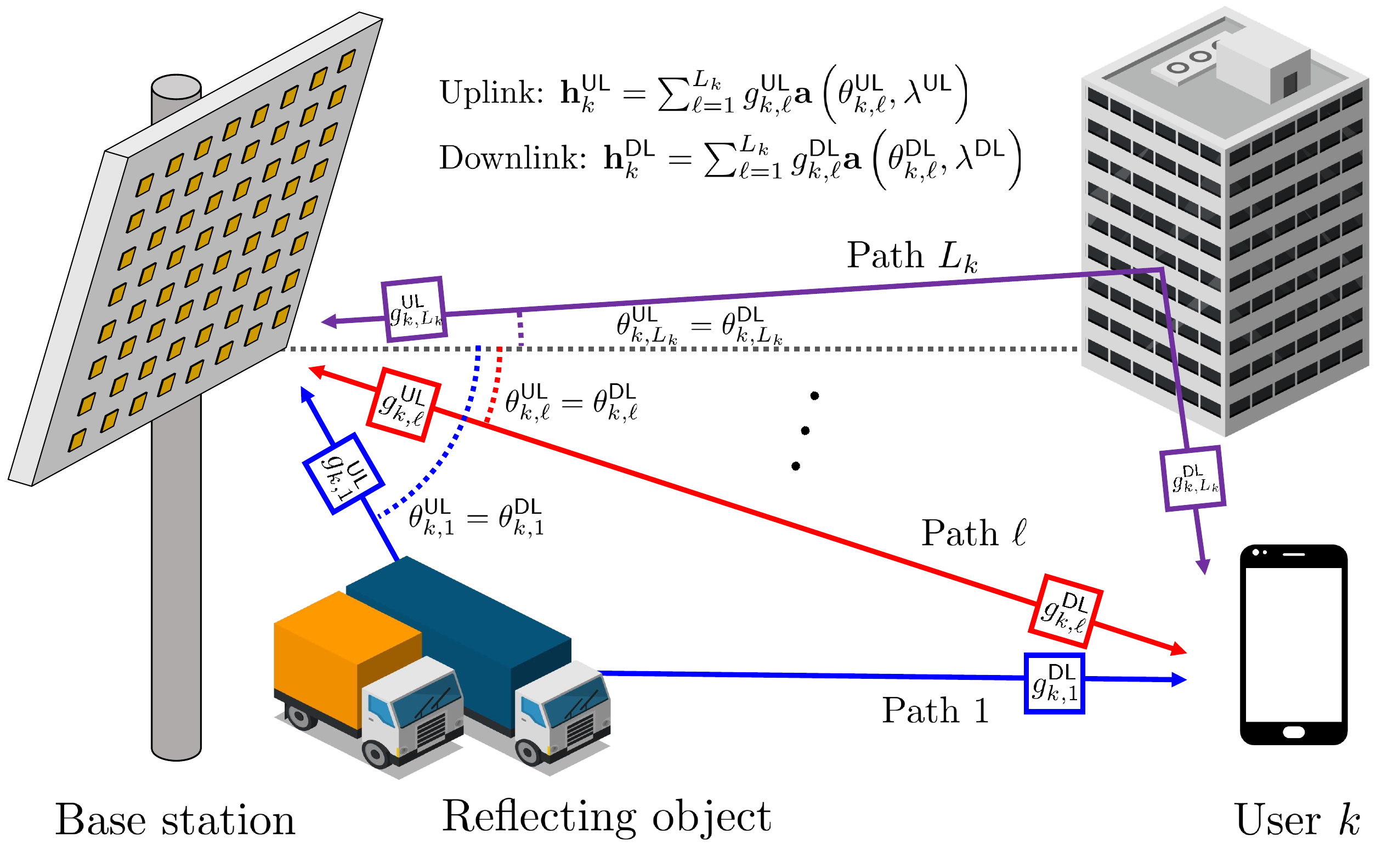}
  \caption{The UL and DL multipath channel models of FDD massive MIMO system.}\label{fig: figures_FDD_MUMIMO}
\end{figure}

\subsection{UL and DL Channel Models}
\subsubsection{UL channel model} 
We denote the UL channel from the user $k \in \{1, \cdots, K\}$ to the BS by $\mathbf{h}_k^{\sf UL} \in \mathbb{C}^N$. This UL channel vector is modeled as a weighted sum of array response vectors. When the uniform linear array (ULA) antenna elements are adopted at the BS, the narrowband UL channel model can be written as in \cite{tse2005fundamentals}: 
\begin{align}\label{eq: UL channel}
    \mathbf{h}_k^{\sf UL} = \sum_{\ell = 1}^{L_k^{\sf UL}} g_{k,\ell}^{\sf UL} \mathbf{a}\left(\theta_{k,\ell}^{\sf UL}, \lambda^{\sf UL}\right),
\end{align}
where $L_k^{\sf UL} \in \mathbb{Z}^+$ and $g_{k,\ell}^{\sf UL} \in \mathbb{C}$ denote the number of channel paths and the $\ell$th complex channel gain of the UL channel for user $k$. The array response vector is a function of AoAs $\theta_{k,\ell}^{\sf UL}$ and wavelength $\lambda^{\sf UL}$, which is defined as 
\begin{align}
    &\mathbf{a}\left(\theta_{k,\ell}^{\sf UL}, \lambda^{\sf UL}\right)  \nonumber \\
    &\ \ = \left[1, e^{-j\frac{2\pi}{\lambda^{\sf UL}}d\sin\theta_{k,\ell}^{\sf UL}}, \cdots, e^{-j\frac{2\pi}{\lambda^{\sf UL}}(N-1)d\sin\theta_{k,\ell}^{\sf UL}} \right]^\top,
\end{align}
where $d$ denote the inter-antenna spacing. The $\ell$th complex channel gain is also given by
\begin{align}\label{eq: g_UL}
    g_{k,\ell}^{\sf UL} = \beta_{k,\ell}^{\sf UL} e^{-j\frac{2\pi}{\lambda^{\sf UL}}r_{k,\ell}^{\sf UL} + \phi_{k,\ell}^{\sf UL}  },
\end{align}
where $\beta_{k,\ell}^{\sf UL} \in \mathbb{R}^+$ is the $\ell$th path loss of the UL channel of user $k$, $r_{k,\ell}^{\sf UL} \in \mathbb{R}^+$ is the signal traversing distance of the $\ell$th path of user $k$, and $\phi_{k,\ell}^{\sf UL} \in [0, 2\pi)$ is a random phase rotation by path reflections in the $\ell$th UL path of user $k$.

\subsubsection{DL channel model} 
Analogous to the UL channel model in (\ref{eq: UL channel}), the DL channel from the BS to user $k$ is represented as the linear combination of the array response vectors $\mathbf{a}(\theta_{k,\ell}^{\sf DL}, \lambda^{\sf DL})$ with the DL channel gain weights $g_{k,\ell}^{\sf DL}$ as 
\begin{align}\label{eq: DL channel}
    \mathbf{h}_k^{\sf DL} = \sum_{\ell = 1}^{L_k^{\sf DL}} g_{k,\ell}^{\sf DL} \mathbf{a}\left(\theta_{k,\ell}^{\sf DL}, \lambda^{\sf DL}\right),  
\end{align}
where $L_k^{\sf DL}$ is the number of channel paths for the DL channel. Similarly, the DL array response vector is written as
\begin{align}
    &\mathbf{a}\left(\theta_{k,\ell}^{\sf DL}, \lambda^{\sf DL}\right)  \nonumber \\
    &\ \ = \left[1, e^{-j\frac{2\pi}{\lambda^{\sf DL}}d\sin\theta_{k,\ell}^{\sf DL}}, \cdots, e^{-j\frac{2\pi}{\lambda^{\sf DL}}(N-1)d\sin\theta_{k,\ell}^{\sf DL}} \right]^\top,
\end{align}
where $\lambda^{\sf DL}$ is the wavelength of the DL signal. The DL channel gain is defined as
\begin{align}\label{eq: g_DL}
    g_{k,\ell}^{\sf DL} = \beta_{k,\ell}^{\sf DL} e^{-j\frac{2\pi}{\lambda^{\sf DL}}r_{k,\ell}^{\sf DL} + \phi_{k,\ell}^{\sf DL}  },
\end{align}
where $\beta_{k,\ell}^{\sf DL} \in \mathbb{R}^+$is the $\ell$th path loss of the DL channel from the BS to user $k$, $r_{k,\ell}^{\sf DL} \in \mathbb{R}^+$ is the DL signal traversing distance of the $\ell$th path of user $k$, and $\phi_{k,\ell}^{\sf DL} \in [0, 2\pi)$ is a random phase rotation by path reflections in the $\ell$th DL path of user $k$.


\subsection{ Frequency-Invariant Channel Parameters}

From the UL and DL channels in (\ref{eq: UL channel}) and (\ref{eq: DL channel}), we can easily see that the UL and DL channels are not reciprocal, i.e., $ \mathbf{h}_k^{\sf UL}\neq  \mathbf{h}_k^{\sf DL}$. Nevertheless, we can identify some frequency-invariant parameters in the UL and DL channels \cite{3GPP25996,tse2005fundamentals}:

\begin{itemize}
\item The AoAs and the AoDs, $\theta_{k,\ell}^{\sf UL}$ and $\theta_{k,\ell}^{\sf DL}$, are reciprocal regardless of carrier frequencies due to the geometric symmetry of the antennas and the signal traversing paths \cite{dai2018fdd,ding2018dictionary}. 
     
    
\item  The channel path gain in UL and DL, $|g_{k,\ell}^{\sf UL}|=\beta_{k,\ell}^{\sf UL}$ and $|g_{k,\ell}^{\sf DL}|=\beta_{k,\ell}^{\sf DL}$ are quasi-reciprocal, i.e., $|g_{k,\ell}^{\sf UL}|\simeq |g_{k,\ell}^{\sf DL}|$. The reason is that the UL and DL signal traversing distance, $r_{k,\ell}^{\sf UL}$ and $r_{k,\ell}^{\sf DL}$, are identical due to the geometric symmetry; thereby, they share the quasi-identical path loss as long as the discrepancy between the UL and DL carrier frequencies are not significant. See more validation of this assumption in \cite{han2023fdd,3GPP25996,tse2005fundamentals}.
    
\end{itemize} 
For the convenience of a mathematical expression, using the UL and DL channel reciprocity, we simplify AoAs and AoDs by omitting the superscripts $\theta_{k,\ell} = \theta_{k,\ell}^{\sf UL} = \theta_{k,\ell}^{\sf DL}$. Similarly, we simplify the signal traversing distance, path attenuation, and the number of channel paths as $r_{k,\ell}=r_{k,\ell}^{\sf UL}=r_{k,\ell}^{\sf DL}$, $\beta_{k,\ell}=\beta_{k,\ell}^{\sf UL}=\beta_{k,\ell}^{\sf DL}$,  and $L_k = L_k^{\sf UL} = L_k^{\sf DL}$. 

\subsection{DL Channel Parameter Acquisition}
We present the DL channel reconstruction method, which integrates both frequency-invariant parameters—namely, the  AoAs and the channel path gains extracted from UL pilots—and \(B_{k,\ell}\)-bit DL CSI feedback per channel path. 

\vspace{0.1cm}
{\bf Acquisition of the AoAs and path gains from the UL pilot}: We assume that the BS has perfect knowledge of DL AoAs, $\theta_{k,\ell}$, and the channel path gains $\beta_{k,\ell}$ for each user \(k\) within the set \([K]\) and each path \(\ell\) within the set \([L_k]\), derived from UL pilot signals. Although these assumptions are somewhat unrealistic, these parameters can be precisely estimated using the classical direction of arrival estimation methods, such as MUSIC \cite{MUSIC}, root-MUSIC \cite{root-music}, and ESPRIT \cite{ESPRIT} along with spatial smoothing techniques \cite{Spatial}, provided the number of UL pilot signals significantly exceeds the number of antennas. When the number of UL pilots is limited, such parameter estimation can be erroneous. To illustrate the impact of estimation accuracy, we will present the DL sum spectral efficiency performance under conditions of inaccurate parameter estimation in Section \ref{Sec: Simulation Results}. 

\vspace{0.1cm}
{\bf Acquisition of the path phase information from limited feedback}: We assume that user \(k \in [K]\) can estimate the multipath channel coefficient \(g^{\text{DL}}_{k,\ell}\) of the DL channel \(\mathbf{h}_k^{\text{DL}}\) from the DL pilots. This assumption is feasible in practice when using orthogonal frequency division multiplexing (OFDM) DL transmissions. Specifically, each user estimates the time-domain multipath channel components from the DL pilots assigned over multiple subcarriers. It can then construct the narrowband DL channel coefficients \(g^{\text{DL}}_{k,\ell}\) for a particular subcarrier using linear interpolation based on the discrete-time Fourier transform \cite{OFDM}.

After estimating the multipath channel coefficient \(g^{\text{DL}}_{k,\ell}\) of the $\ell$th channel path, user $k\in[K]$ quantizes the phase information of the channel path, \(\angle g^{\text{DL}}_{k,\ell}\), with a $B_{k,\ell}$-bit uniform scalar quantization and provides feedback to the BS via the error-free feedback link. Let 
\begin{align}
    \mathcal{C}_{k,\ell} = \left\{0, \frac{2\pi}{2^{B_{k,\ell}}}, \cdots,  \frac{2\pi}{2^{B_{k,\ell}}}\left( 2^{B_{k,\ell}} - 1\right) \right\}
\end{align}
 be the codebook of the uniform phase quantization. Then, the quantization function $Q(\cdot):[0,2\pi]\rightarrow \mathcal{C}_{k,\ell}$ produces the quantization output for the phase information as
\begin{align}
    Q_{B_{k,\ell}}(\angle g^{\sf DL}_{k,\ell}) = q^{\sf DL}_{k,\ell}
\end{align}
 with
\begin{align}
    q^{\sf DL}_{k,\ell} = \underset{c_{j,k,\ell} \in \mathcal{C}_{k,\ell} }{\rm argmin}\left\vert \angle g^{\sf DL}_{k,\ell}- c_{j,k,\ell} \right\vert,
\end{align}
where $c_{j,k,\ell}=\frac{2\pi (j-1)}{2^{B_{k,\ell}}}$. Then, the feedback error for the $l$th channel path phase of user $k$ is defined as
\begin{align}
    \delta_{k,\ell} =  \angle g^{\sf DL}_{k,\ell} -q^{\sf DL}_{k,\ell}, \label{eq:quanerror}
\end{align}
where $\delta_{k,\ell}$ is uniformly distributed in $ \left[- \frac{2\pi}{2^{B_{k,\ell}+1}}, \frac{2\pi}{2^{B_{k,\ell}+1}}\right]$.

\subsection{DL Channel Reconstruction Problem}

By integrating the path gains $\{\beta_{k,1},\ldots, \beta_{k,L_k}\} \in \mathbb{R}^{L_k}$, the AoAs $\{\theta_{k,1},\ldots, \theta_{k,L_k}\} \in [0,2\pi]^{L_k}$ and the quantized phase information $\{q^{\sf DL}_{k,1},\ldots, q^{\sf DL}_{k,L_k}\} \in [0,2\pi]^{L_k}$, the BS reconstructs the DL channels. Let $g:\left(\mathbb{R}^{L_k}, [0,2\pi]^{L_k},[0,2\pi]^{L_k} \right)\rightarrow \mathbb{C}^{N}$ be a reconstruction function that maps the input channel parameters to the DL channel as
\begin{align}
    \mathbf{\hat h}_k^{\sf DL}=g\left(\{\beta_{k,\ell}\}_{\ell=1}^{L_k},\{\theta_{k,\ell}\}_{\ell=1}^{L_k} ,\{q^{\sf DL}_{k,\ell}\}_{\ell=1}^{L_k}  \right).
\end{align}
Our goal is to identify the MSE-optimal DL channel reconstruction function such that
\begin{align}
    \mathbf{\hat h}_k^{\sf DL,MMSE}=  \underset{g(\cdot)}{\rm argmin} \ \   \mathbb{E}\left[\left( \mathbf{ h}_k^{\sf DL} - \mathbf{\hat h}_k^{\sf DL}\right)^2\right], \label{eq:MSE}
\end{align}
where the expectation is taken over the quantization errors of the phase information $\delta_{k,\ell}$, which is uniformly distributed in $ \left[- \frac{2\pi}{2^{B_{k,\ell}+1}}, \frac{2\pi}{2^{B_{k,\ell}+1}}\right]$.



\section{Bayesian Approach to DL Channel Reconstruction}\label{Sec: DL Reconst}
In this section, we first present the MSE-optimal DL channel reconstruction method that minimizes the MSE in \eqref{eq:MSE}. Then, we establish the minimum MSE (MMSE) function in terms of the number of CSI feedback bits $B_{k,\ell}$ for $\ell\in [L_k]$.

\begin{figure}[t]
	\centering
    \includegraphics[width=\linewidth]{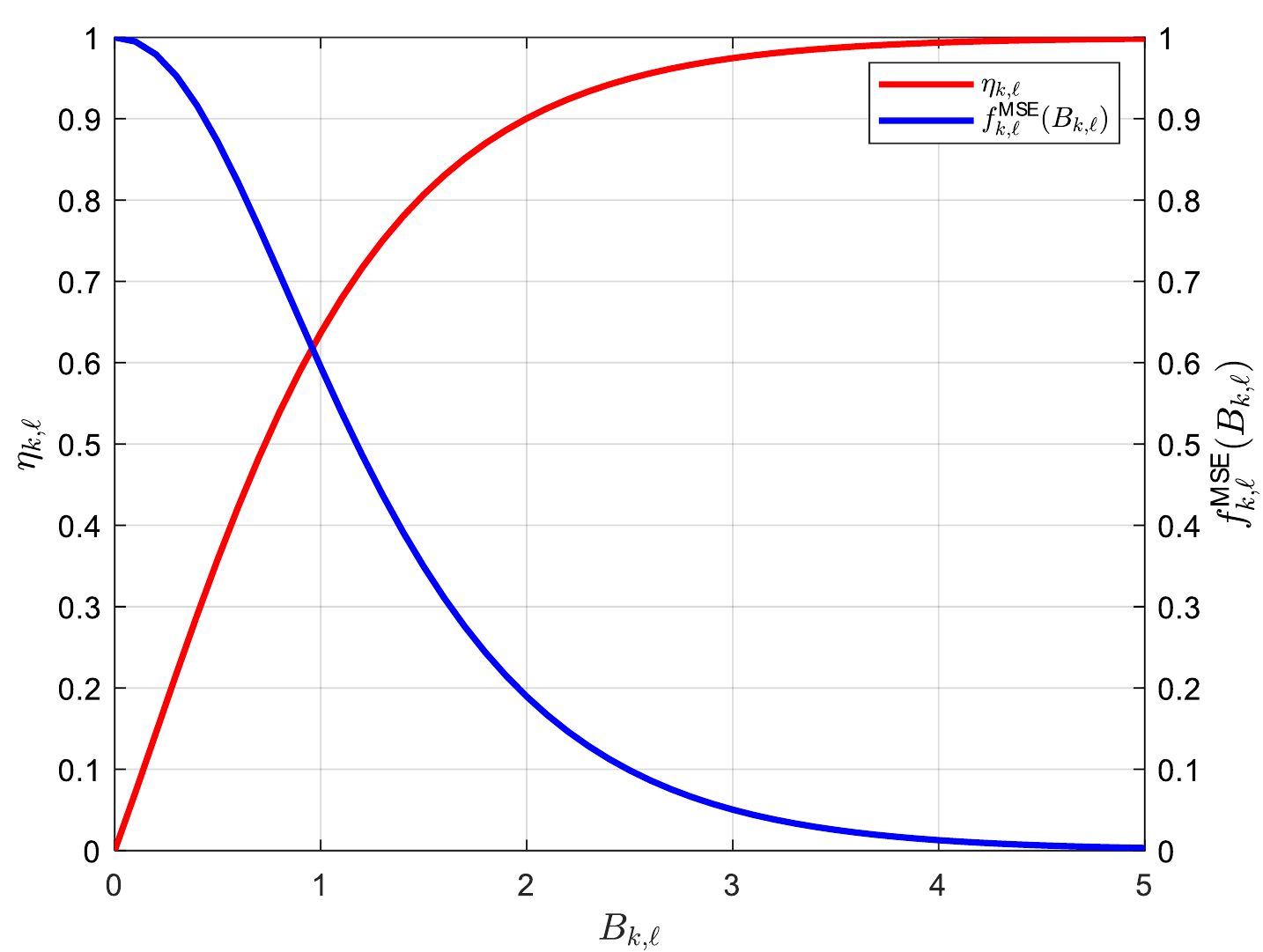}
  \caption{The phase error compensation parameter  $\eta_{k,\ell}$ and the NMMSE $f_{k,\ell}^{\sf MSE}(B_{k,\ell})$ as increasing  $B_{k,\ell}$.} \label{fig: function_eta}
\end{figure}

\subsection{MSE-Optimal DL Channel Reconstruction}
The following theorem is the main result of this paper. 
 
\begin{thm}\label{thm: MSE-OptDLCh}
     The MSE-optimal solution for the optimization problem in \eqref{eq:MSE} is given by
    \begin{align}
          \hat{\mathbf{h}}_k^{\sf DL, MMSE}=\sum_{\ell=1}^{L_k} \eta_{k,\ell}\beta_{k,\ell}e^{jq_{k,\ell}^{\sf DL}} \mathbf{a}\left(\theta_{k,\ell}, \lambda^{\sf DL}\right), \label{eq: MMSE DL_main}
    \end{align}
    where $ \eta_{k,\ell}$ is the phase quantization error compensation parameter defined as
    \begin{align}
        \eta_{k,\ell} = \frac{2^{B_{k,\ell}}}{\pi} \sin \left( \frac{\pi}{2^{B_{k,\ell}}}\right).\label{eq: eta_main}
    \end{align}
\end{thm}

\begin{IEEEproof}
    See Appendix \ref{Prf: MSE-OptDLCh}.
\end{IEEEproof}
\vspace{0.1cm}
 
To understand the MSE-optimal DL channel reconstruction solution presented in Theorem \ref{thm: MSE-OptDLCh}, it is instructive to compare it with the true DL channel \(\mathbf{h}_k^{\sf DL}\) as described in \eqref{eq: DL channel}. Assuming the BS is aware of the AoAs, \(\theta_{k,\ell}\), and the path gains, \(\beta_{k,\ell}\), the differences between \(\mathbf{h}_k^{\sf DL}\) and \(\hat{\mathbf{h}}_k^{\sf DL, MMSE}\) occur in two main aspects: 1) the phase quantization term \(e^{j q_{k,\ell}^{\sf DL}} = e^{j (\angle g_{k,\ell}^{\sf DL} + \delta_{k,\ell})}\) and 2) the phase error compensation by \(\eta_{k,\ell}\). As the feedback bit \(B_{k,\ell}\) increases, it becomes clear that the phase quantization error, as indicated in \eqref{eq:quanerror}, diminishes to zero, namely, 
\begin{align}
    \lim_{B_{k,\ell}\rightarrow \infty} \delta_{k,\ell} \rightarrow 0.
\end{align}
Similarly, Fig. \ref{fig: function_eta} shows that the phase error compensation term in \eqref{eq: eta_main} also approaches to one, i.e., 
\begin{align}
    \lim_{B_{k,\ell}\rightarrow \infty} \eta_{k,\ell} \rightarrow 1.
\end{align}
As a result, the proposed MSE-optimal DL channel reconstruction method can asymptotically recover the true DL channel with an increasing number of feedback bits, i.e.,
\begin{align}
     \lim_{B_{k,1},\ldots, B_{k,L_k}\rightarrow \infty}   \hat{\mathbf{h}}_k^{\sf DL, MMSE} \rightarrow  {\mathbf{h}}_k^{\sf DL}.
\end{align}
 
We note that the proposed MSE-optimal DL reconstruction method does not depend on the UL carrier frequency. In addition, this observation differs from the prior study in \cite{han2023fdd}, where the channel phase terms for path reflections in UL and DL, \(\phi_{k,\ell}^{\sf UL}\) and \(\phi_{k,\ell}^{\sf DL}\), are assumed to be identical according to the standard channel model \cite{3GPP25996, vasisht2016eliminating, han2023fdd}. However, in this paper, we generalize this assumption to make it more practical by modeling these phase terms, \(\phi_{k,\ell}^{\sf UL}\) and \(\phi_{k,\ell}^{\sf DL}\), as statistically independent.


In addition, when the MUSIC algorithm is applied along with the spatial smoothing technique \cite{MUSIC,Spatial}, the AoAs and the path gains can be estimated with some errors. Let \({\hat \theta}_{k,\ell}\) and \({\hat \beta}_{k,\ell}\) be the estimates of the AoAs and path gains, respectively. Then, the DL channel reconstruction method can be readily modified as follows:   
\begin{align}
         \hat{\mathbf{h}}_k^{\sf DL, MMSE}=\sum_{\ell=1}^{L_k} \eta_{k,\ell}{\hat \beta}_{k,\ell}e^{jq_{k,\ell}^{\sf DL}} \mathbf{a}\left({\hat \theta}_{k,\ell}, \lambda^{\sf DL}\right). \label{eq: MMSE DL_main2}
   \end{align}
The effect of the AoA and the path gain estimation errors will be discussed in Section \ref{Sec: Simulation Results}.

\subsection{MSE Analysis}

We establish the MSE for the DL channel reconstruction when applying the proposed DL channel reconstruction method in (\ref{eq: MMSE DL_main}). The MMSE is stated in the following theorem.

\begin{thm}\label{thm:MSE}
    The MMSE function is represented in terms of CSI feedback bits $B_{k,\ell}$ for $\ell\in [L_k]$ as
    \begin{align}
        &\mathbb{E}\left[\!\left( \mathbf{ h}_k^{\sf DL} -\hat{\mathbf{h}}_k^{\sf DL, MMSE}\right)^2\right] \nonumber \\
        & \hspace{1cm}= \sum_{\ell=1}^{L_k}\!N \beta_{k,\ell}^2 \!\left(\!1 \!-\! \left(\frac{2^{B_{k,\ell}}}{\pi} \sin \left(  \frac{\pi}{2^{B_{k,\ell}}}\!   \right)\! \right)^2 \right). \label{eq: MSE_k}
    \end{align}    
\end{thm}
\begin{IEEEproof}
    See Appendix \ref{Prf: MSE}.
\end{IEEEproof}
\vspace{0.1cm}

Theorem \ref{thm:MSE} clearly shows how the MMSE value can vary according to the number of bits allocated for each channel path $B_{k,\ell}$. To illustrate this, we define the normalized MMSE (NMMSE) for the $\ell$th channel path as
\begin{align}
    f_{k,\ell}^{\sf MSE}(B_{k,\ell}) &=\frac{\mathbb{E}\left[\left( \mathbf{ h}_{k,\ell}^{\sf DL} - \mathbf{\hat h}_{k,\ell}^{\sf DL, MMSE}\right)^2\right]}{N \beta_{k,\ell}^2} \nonumber\\
    &= 1 - \left(\frac{2^{B_{k,\ell}}}{\pi} \sin \left(  \frac{\pi}{2^{B_{k,\ell}}}   \right) \right)^2.
\end{align}
Fig. \ref{fig: function_eta} shows the NMMSE decreasing as the number of feedback bits \(B_{\ell,k}\) increases. As can be seen, the NMMSE diminishes as more CSI feedback bits are used to quantize the phase of the channel, which aligns with our intuition. In addition, when the quantization bits are allocated uniformly across the channel paths, i.e., \(B_{k,\ell} = B_k\) for \(\ell \in [L_K]\), each channel path contributes differently to the overall MMSE as described in \eqref{eq: MSE_k}. Specifically, a path with higher channel power \(\beta_{k,\ell}^2\) significantly increases the MMSE. This observation motivates us to design an optimal feedback bit allocation strategy across the channel path to minimize the MMSE, especially when the total number of feedback bits is limited \(\sum_{\ell=1}^{L_k} B_{k,\ell} \leq B^{\sf tot}\).
 


\section{Optimal Feedback Bit Allocation}\label{Sec: Bit Allo}

\begin{figure}[t]
	\centering
    \includegraphics[width=\linewidth]{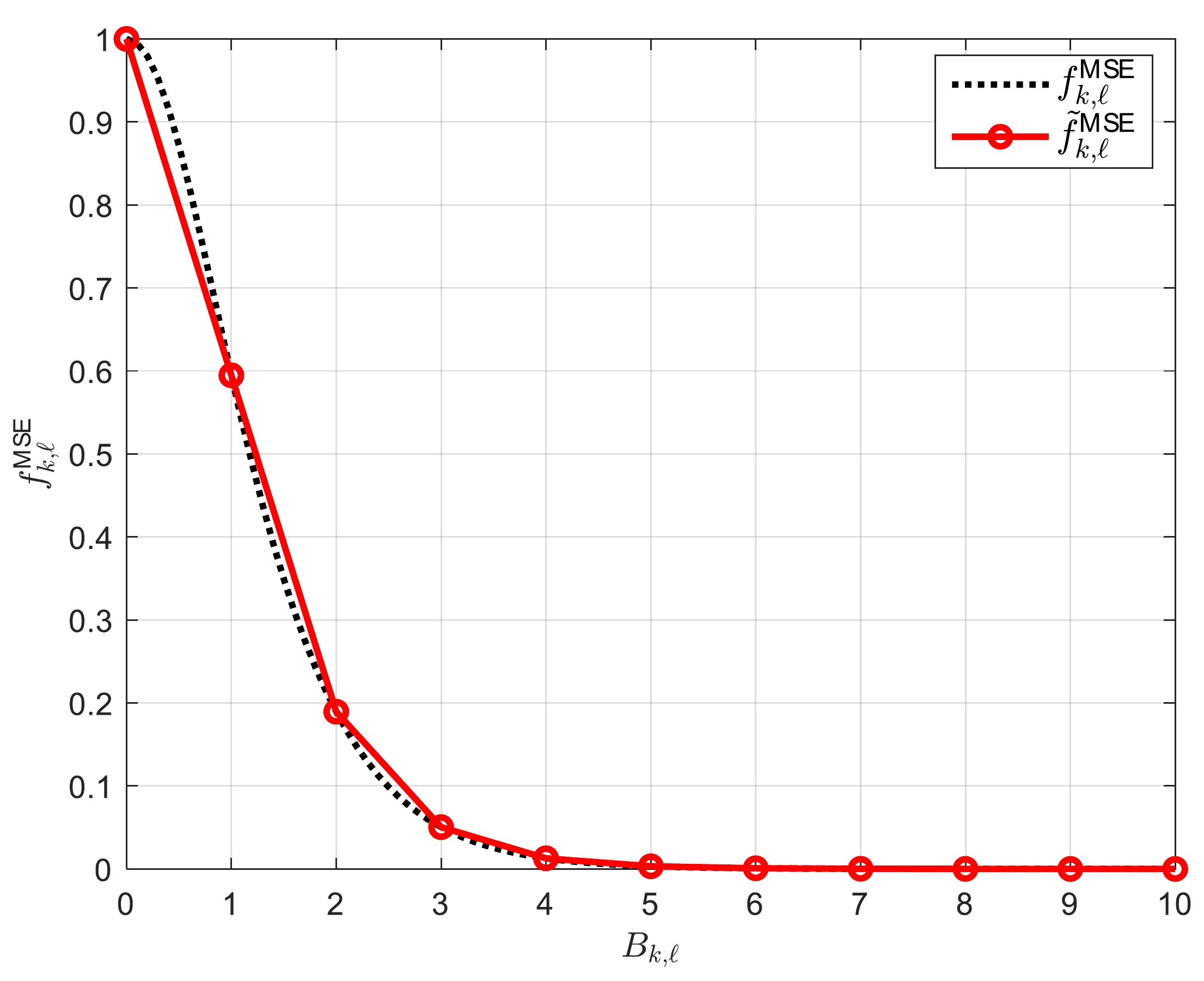}
  \caption{The piecewise linear approximation of $ {f}^{\sf MSE}_{k,\ell}(x)$.}\label{fig: function_MSE_MSEconst}
\end{figure}

In this section, we propose the optimal feedback bit allocation strategy to minimize the MSE for DL channel reconstruction, given a fixed total number of feedback bits. Under the constraint that the sum of feedback bits \(\sum_{\ell=1}^{L_k} B_{k,\ell} = B^{\sf tot}\), the MSE-optimal feedback bit allocation solution is obtained by solving the following optimization problem:
\begin{align}
  \!\!\!  \mathscr{P}1:& \underset{B_{k,1}, \cdots, B_{k,L_k}}{\rm argmin} \sum_{\ell=1}^{L_k} \beta_{k,\ell}^2 f_{k,\ell}^{\sf MSE}(B_{k,\ell}) \label{eq: OPT1}\\
    &\text{subject to  } \sum_{\ell=1}^{L_k} B_{k,\ell} = B^{\sf tot} \\
    &~~~~~~~~~~~~~ B_{k,\ell} \in \mathbb{Z}_{\ge 0}. \label{eq: OPT1_const2}
\end{align}
Unfortunately, this optimization presents a non-convex problem because the NMMSE, \(f^{\sf MSE}_{k,\ell}\), is a non-convex function, as illustrated in Fig. \ref{fig: function_eta}. Additionally, the integer constraint on \(B_{k,\ell} \in \mathbb{Z}_{\ge 0}\) complicates obtaining the optimal bit allocation strategy in closed form. To overcome these challenges, we employ a piecewise linear continuous approximation of the continuous NMMSE function \(f^{\sf MSE}_{k,\ell}\). Specifically, we define a piecewise linear approximation that satisfies the equalities at the feasible integer set, i.e., 
\begin{align}
    {\tilde f}^{\sf MSE}_{k,\ell}(x)=f^{\sf MSE}_{k,\ell}(x)
\end{align}
at $x \in \mathbb{Z}_{\ge 0}$. As illustrated in Fig. \ref{fig: function_MSE_MSEconst}, the piecewise linear approximation $ {\tilde f}^{\sf MSE}_{k,\ell}(x)$ is a function that linearly interpolates the true function values $ { f}^{\sf MSE}_{k,\ell}(x)$ for $x \in \mathbb{Z}_{\ge 0}$. The following lemma shows some important properties of this piecewise linear approximation $ {\tilde f}^{\sf MSE}_{k,\ell}(x)$, which will be used to design the optimal bit allocation strategy. 

\begin{lem}\label{lem1}
The function \( {\tilde f}^{\sf MSE}_{k,\ell}(x)\) is convex and strictly decreasing.\end{lem}

\begin{IEEEproof}
    See Appendix \ref{Prf: convex_dec}.
\end{IEEEproof}

\vspace{0.1cm}

Based on Lemma \ref{lem1}, we can reformulate the optimization problem presented in \eqref{eq: OPT1} as follows:
\begin{align}
    \mathscr{P}1^\prime:& \underset{B_1, \cdots, B_{k,L_k}}{\rm argmin} \sum_{\ell=1}^{L_k} \beta_{k,\ell}^2\ \tilde{f}^{\sf MSE}_{k,\ell} \left(B_{k,\ell} \right)  \label{eq: OPT2}\\
    &\text{subject to  } \sum_{\ell=1}^{L_k} B_{k,\ell} = B^{\sf tot} \\
    &\text{subject to  }  B_{k,\ell} \in \mathbb{Z}_{\ge 0}. \label{eq: OPT2_const2}
\end{align}
Although the objective function in this reformulated optimization problem is convex, the presence of an integer constraint on $ B_{k,\ell}$ complicates the identification of the optimal feedback bit allocation method. Nevertheless, the optimal solution can be achieved by employing classical discrete optimization techniques, as in \cite{fox1966discrete}. The convexity and strictly decreasing nature of the objective function enable the incremental bit allocation scheme to efficiently identify the optimal solution through marginal analysis. The theorem below articulates this assertion more formally.



\begin{figure}[t]
	\centering
    \includegraphics[width=\linewidth]{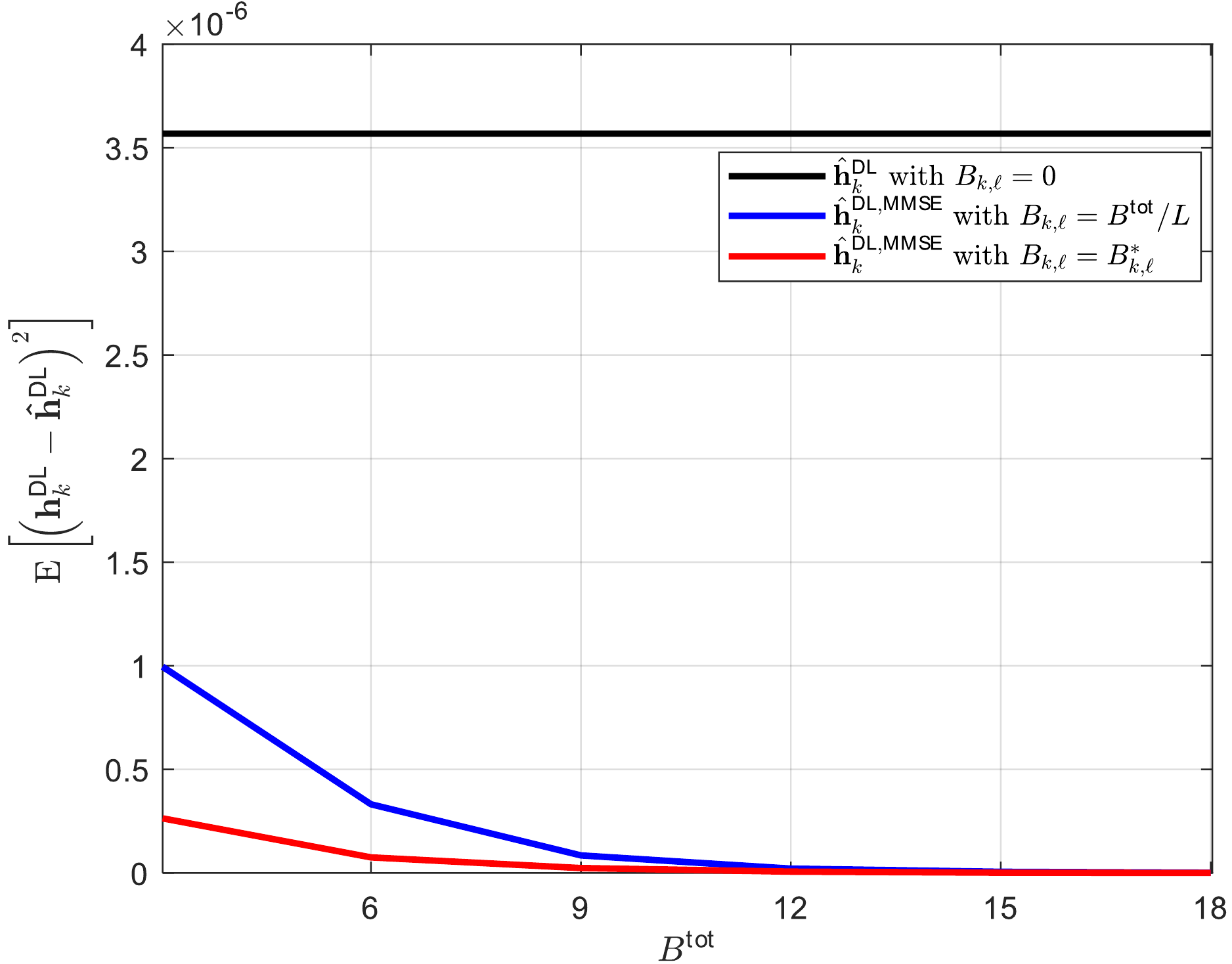}
  \caption{The MMSE comparison according to the feedback bit allocation strategy. We set the simulation parameters as $(L_k,K,N) = (3, 16,256)$.}\label{fig: results_MSE_comparison}
\end{figure}

\begin{thm}\label{thm: bit_allo_optimality}
    If ${\tilde f}^{\sf MSE}_{k,\ell} (B_{k,\ell})$ is convex and strictly decreasing, the bit allocation $B_{k,\ell}^*$ generated by Algorithm \ref{Alg: Opt Bit Allo} is optimal for $\mathscr{P}1^\prime$.
\end{thm}
\begin{IEEEproof}
    See Appendix \ref{Prf: bit_allo_optimality}.
\end{IEEEproof}
\vspace{0.2cm}

To validate the effectiveness of the proposed bit allocation method, we compared its MMSE performance with that of a uniform bit allocation strategy. As a benchmark, we also draw the MMSE value when constructing a DL channel with only UL pilots, i.e., no CSI feedback is used. As illustrated in Fig. \ref{fig: results_MSE_comparison}, the proposed strategy consistently outperforms the uniform approach across all feedback bit constraints where \(B^{\sf tot} > 0\). Notably, the MMSE improvement with the proposed method is more significant when the total feedback bits are limited. This demonstrates that the choice of bit allocation strategy is crucial under stringent feedback bit constraints. However, as the total number of feedback bits increases, the sensitivity to the bit allocation strategy decreases, and the uniform bit allocation method becomes optimal.

\begin{algorithm}[t]
    \caption{Optimal Bit Allocation Procedure}
    Initialization: $t=0,~ B_{k,\ell}^{(0)} = 0$\;
    \While{$\sum_{\ell=1}^{L_k} B_{k,\ell}^{(t)} \le B^{\sf tot}$}{ 
    $t \leftarrow t+1 $  \;
    $\ell^* = \underset{\ell=1,\cdots, L_k}{\rm argmax} \left(\beta_{k,\ell}^2\ {\tilde f}^{\sf MSE}_{k,\ell} \left(B_{k,\ell}^{(t-1)} \right) \right.$ \
    \hspace*{2.3cm}$\left.- \beta_{k,\ell}^2\ {\tilde f}^{\sf MSE}_{k,\ell} \left(B_{k,\ell}^{(t-1)}+1 \right) \right)$ \;
    \For{ $\ell = 1, \cdots, L_k$}{
    \eIf{$\ell = \ell^*$}
    {$B_{k,\ell}^{(t)} = B_{k,\ell}^{(t-1)} + 1 $ \;}
    {$B_{k,\ell}^{(t)} = B_{k,\ell}^{(t-1)}  $ \;}
    }
    }
    $B_{k,\ell}^* = B_{k,\ell}^{(t)}~~ \forall \ell$\;
    \label{Alg: Opt Bit Allo}
\end{algorithm}

\section{Robust DL Massive MIMO Precoding}\label{Sec: Robust DL Precoding}
In this section, we present a robust massive MIMO precoding method that maximizes sum-spectral efficiency. The key idea behind the proposed robust MIMO precoding method is to utilize both the MSE-optimal DL channel estimate \(\hat{\mathbf{h}}_k^{\sf DL, MMSE}\) and the corresponding channel reconstruction error covariance matrix.

\subsection{Error Covariance Matrix for \(\hat{\mathbf{h}}_k^{\sf DL, MMSE}\)}

The following lemma establishes the error covariance matrix for \(\hat{\mathbf{h}}_k^{\sf DL, MMSE}\).


\begin{lem}\label{lem2}
    Let $\mathbf{e}_k = \mathbf{h}_k^{\sf DL} - \hat{\mathbf{h}}_k^{\sf DL, MMSE}$ be the DL channel reconstruction error. The error covariance matrix with knowledge of $\{\mathbf{h}_k^{\sf UL}, \{\beta_{k,\ell}\}_{\ell=1}^{L_k},\{\theta_{k,\ell}\}_{\ell=1}^{L_k} ,\{q^{\sf DL}_{k,\ell}\}_{\ell=1}^{L_k}\}$ for $k \in \{1, \cdots, K\}$ is
    \begin{align}
        \mathbf{\Phi}_k^{\sf MMSE} &= \mathbb{E} \left[ \mathbf{e}_k  \mathbf{e}_k ^{\sf H} |\mathbf{h}_k^{\sf UL}, \{\beta_{k,\ell}\}_{\ell=1}^{L_k},\{\theta_{k,\ell}\}_{\ell=1}^{L_k} ,\{q^{\sf DL}_{k,\ell}\}_{\ell=1}^{L_k}\right] \nonumber \\
        &= \sum_{\ell=1}^{L_k} \beta_{k,\ell}^2 (1-\eta_{k,\ell}^2) \mathbf{a}\left(\theta_{k,\ell}, \lambda^{\sf DL}\right)\mathbf{a}\left(\theta_{k,\ell}, \lambda^{\sf DL}\right)^{\sf H}.
    \end{align}


    
\end{lem}


\begin{IEEEproof}
    See Appendix \ref{Prf: Error cov}.
\end{IEEEproof}
\vspace{0.2cm}

Lemma \ref{lem2} clearly shows that the MSE matrix approach the zero matrix as $\lim_{B_{k,\ell}\rightarrow \infty}\eta_{k,\ell}=1$. Next, we provide the error analysis between $\mathbf{h}_k^{\sf DL} (\mathbf{h}_k^{\sf DL})^{\sf H}$ and $\hat{\mathbf{h}}_k^{\sf DL}(\hat{\mathbf{h}}_k^{\sf DL})^{\sf H} + \mathbf{\Phi}_k^{\sf} $, which plays a important role for the robust massive MIMO optimization.

\begin{lem}\label{lem3}
    When the outer product of the DL channel $\mathbf{h}_k^{\sf DL} (\mathbf{h}_k^{\sf DL})^{\sf H}$ is approximated by $\hat{\mathbf{h}}_k^{\sf DL}(\hat{\mathbf{h}}_k^{\sf DL})^{\sf H} + \mathbf{\Phi}_k^{\sf} $, the approximation error matrix $\mathbf{\Delta}_k$ is given by
    \begin{align}
        \mathbf{\Delta}_k = \mathbf{h}_k^{\sf DL} (\mathbf{h}_k^{\sf DL})^{\sf H} -\left( \hat{\mathbf{h}}_k^{\sf DL}(\hat{\mathbf{h}}_k^{\sf DL})^{\sf H} + \mathbf{\Phi}_k^{\sf}\right).
    \end{align}
    Then, the asymptotic error in the number of antennas when $\hat{\mathbf{h}}_k^{\sf DL} = \hat{\mathbf{h}}_k^{\sf DL, MMSE}$ is 
    \begin{align}
        \lim_{N\rightarrow \infty} \frac{1}{N^2}\Vert \mathbf{\Delta}_k\Vert_F^2 =  \sum_{\ell \ne \ell'} 2\left(1 + |\eta_{k,\ell}\eta_{k,\ell'}|^2\right)|\beta_{k,\ell}\beta_{k,\ell'}|^2 \nonumber \\ 
        -4|\eta_{k,\ell}\eta_{k,\ell'}||\beta_{k,\ell}\beta_{k,\ell'}|^2 {\rm Re} \left( e^{j \delta_{k,\ell}}e^{-j \delta_{k,\ell'}}\right)
    \end{align}
\end{lem}
\begin{IEEEproof}
    See Appendix \ref{Prf: Outer apprx}.
\end{IEEEproof}
\vspace{0.2cm}

\begin{figure}[t]
	\centering
    \includegraphics[width=\linewidth]{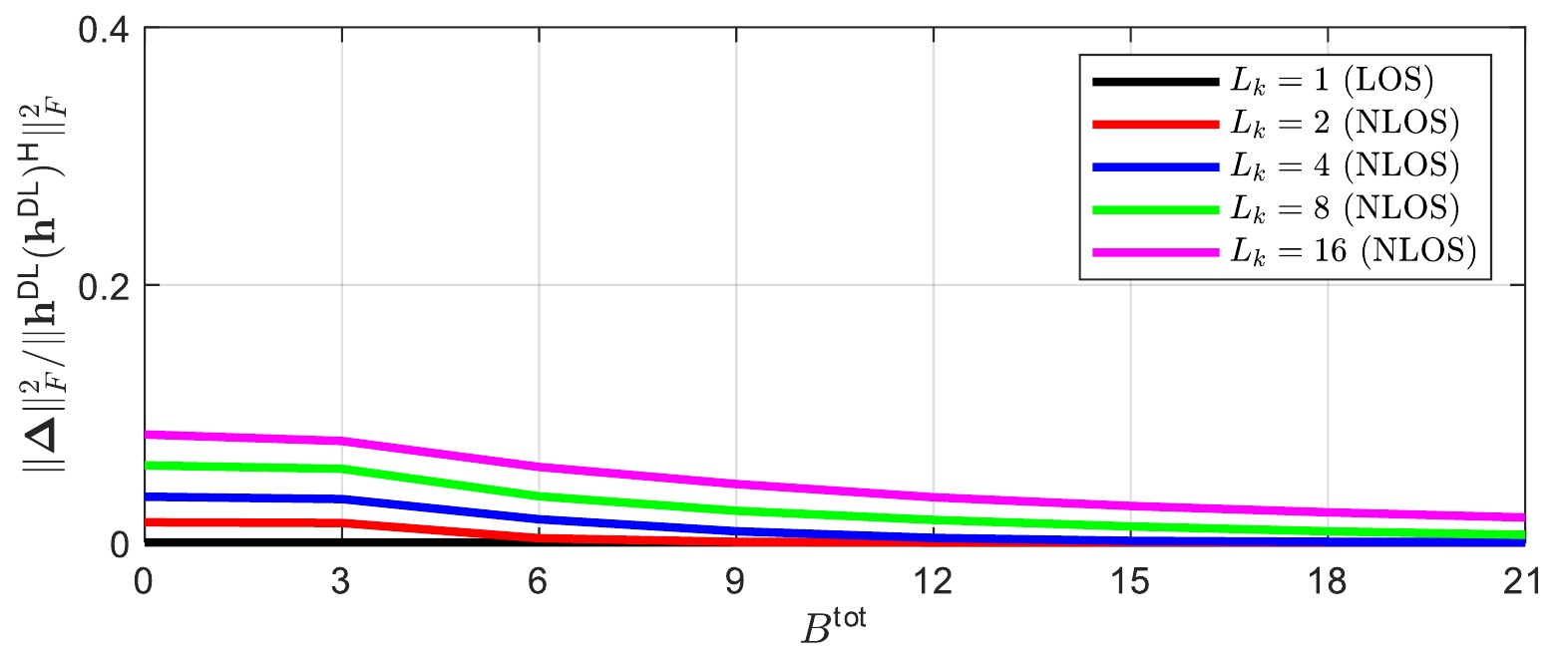}
  \caption{The normalized approximation error with $(K,N) = (16,256)$ and $\frac{\lambda^{\sf UL}}{\lambda^{\sf DL}} = 1.2$ according to the number of feedback bits $B^{\sf tot} \in \{3, 6, 9, \cdots, 21\}$.}\label{fig: results_Delta_comparison}
\end{figure}

From Lemma \ref{lem2} and Lemma \ref{lem3}, we observe that the outer product of the true DL channel \(\mathbf{h}_k^{\sf DL} (\mathbf{h}_k^{\sf DL})^{\sf H}\) can be approximated by \(\hat{\mathbf{h}}_k^{\sf DL}(\hat{\mathbf{h}}_k^{\sf DL})^{\sf H} + \mathbf{\Phi}_k^{\sf}\) in the average sense, i.e.,
\begin{align}
    \mathbb{E}\left[\mathbf{h}_k^{\sf DL} (\mathbf{h}_k^{\sf DL})^{\sf H} \right]  = \mathbb{E} \left[ \hat{\mathbf{h}}_k^{\sf DL}(\hat{\mathbf{h}}_k^{\sf DL})^{\sf H} + \mathbf{\Phi}_k^{\sf}\right], \forall k \in [K].
\end{align}
This claim means that our approximation is unbiased for \(\mathbf{h}_k^{\sf DL} (\mathbf{h}_k^{\sf DL})^{\sf H}\). In addition, the approximation becomes exact when there is only one channel path, i.e., in the case of a line-of-sight (LOS) channel. As shown in Fig. \ref{fig: results_Delta_comparison}, the approximation error becomes more pronounced as the number of channel paths increases. However, the approximation error can become negligible as the total number of feedback bits \(B^{\rm tot}\) increases.


\subsection{Robust Sum Spectral Efficiency Maximization}
We present the sum-spectral efficiency maximization problem when the BS  has imperfect knowledge about the DL channel, i.e., ${\bf \hat h}_{k,\ell}^{\sf DL}$ and the corresponding error covariance matrix $\mathbf{\Phi}_k^{\sf}$. In particular, we take the approach developed in \cite{choi2019joint}, in which a lower bound of the sum-spectral efficiency was characterized when the BS has knowledge of imperfect DL channel from generalized mutual information technique introduced in \cite{GMI}. From \cite{choi2019joint}, under the premise that transmit symbols are drawn from the independent and identical Gaussian distribution with zero mean and variance $P$, the sum-spectral efficiency lower bound is given by
\begin{align}
    &{R}^{\sf LB}(\mathbf{f}_1, \cdots, \mathbf{f}_K) \nonumber \\
    &= \sum_{k=1}^K \log_2\!\left( \!1\!+\! \frac{  \mathbf{f}_k^{\sf H} \left( \hat{\mathbf{h}}_k^{\sf DL}(\hat{\mathbf{h}}_k^{\sf DL})^{\sf H}  + \mathbf{\Phi}_k \right) \mathbf{f}_k  }{\sum_{i\ne k}^K  \mathbf{f}_i^{\sf H}  \left( \hat{\mathbf{h}}_k^{\sf DL}(\hat{\mathbf{h}}_k^{\sf DL})^{\sf H}+ \mathbf{\Phi}_k\right) \mathbf{f}_i \!+\! \frac{\sigma_k^2}{P} } \!\right)\nonumber \\
    &= \log_2\!\left(\! \prod_{k=1}^K \frac{\sum_{i=1}^K  \mathbf{f}_i^{\sf H} \left( \hat{\mathbf{h}}_k^{\sf DL}(\hat{\mathbf{h}}_k^{\sf DL})^{\sf H} + \mathbf{\Phi}_k^{\sf}\right)\mathbf{f}_i + \frac{\sigma_k^2}{P} }{\sum_{i\ne k}^K  \mathbf{f}_i^{\sf H} \left( \hat{\mathbf{h}}_k^{\sf DL}(\hat{\mathbf{h}}_k^{\sf DL})^{\sf H} + \mathbf{\Phi}_k^{\sf}\right)\mathbf{f}_i + \frac{\sigma_k^2}{P} } \!\right)\!\!, \label{eq: sum_spectral_eff_app}
\end{align}
where ${\bf f}_i \in \mathbb{C}^{N\times 1}$ denotes precoding vector carrying data symbol for user $i\in [K]$ and $\sigma_k^2$ is the noise power at the DL receiver of user $k\in [K]$. Unfortunately, this maximization of the sum-spectral efficiency is a well-known NP-hard problem \cite{luo2008dynamic}. While the WMMSE algorithm has been widely employed to address this optimization \cite{christensen2008weighted}, recent findings suggest that the generalized power iteration precoding (GPIP) algorithm is a more efficient solution for joint user selection, beamforming, and power allocation optimization in MU-MIMO systems, regardless of the number of users and antennas \cite{Lee1,choi2019joint, han2020distributed, han2021sparse, han2023fdd, RS-GPIP, GPIP2}. In this paper, we shall develop robust DL precoding using the GPIP algorithm.

\subsection{Robust Massive MIMO Precoding via GPIP}
 
The key idea of the GPIP introduced in \cite{choi2019joint} involves the joint optimization of the precoding vectors \(\{\mathbf{f}_1, \cdots, \mathbf{f}_K\}\) to manage joint power allocation, user selection, and beam design. To accomplish this, we define a unified precoding vector \(\mathbf{f}\) by concatenating all the precoding vectors as 
\begin{align}
    \mathbf{f} = \left[ \mathbf{f}_1^{\sf H}, \cdots , \mathbf{f}_K^{\sf H}\right]^{\sf H} \in \mathbb{C}^{NK\times 1}.
\end{align}
Using this unified precoding vector, the optimization problem aimed at maximizing the lower bound of the sum-spectral efficiency, as defined in \eqref{eq: sum_spectral_eff_app}, under the sum power constraint is formulated as 
\begin{align}
    \mathscr{P}2:~~& \underset{\mathbf{f} \in \mathbb{C}^{NK\times 1}}{\rm argmax}  \prod_{k=1}^K \frac{\mathbf{f}^{\sf H}\mathbf{A}_k\mathbf{f}}{\mathbf{f}^{\sf H}\mathbf{B}_k\mathbf{f}},     \\
    &\text{subject to  }  \Vert \mathbf{f}\Vert_2^2 = 1 ,
\end{align}
where $\mathbf{A}_k \in \mathbb{C}^{NK\times NK}$ and $\mathbf{B}_k \in \mathbb{C}^{NK\times NK}$ are positive semi-definite block diagonal matrix, which are defined as
\begin{align}
    \mathbf{A}_k = \mathbf{I}_K \otimes \left( \hat{\mathbf{h}}_k^{\sf DL}(\hat{\mathbf{h}}_k^{\sf DL})^{\sf H} + \mathbf{\Phi}_k^{\sf}\right) + \frac{\sigma_k^2}{P}\mathbf{I}_{NK},
\end{align}
and
\begin{align}
    \mathbf{B}_k  = \mathbf{A}_k - \mathbf{1}_k\mathbf{1}_k^\top \otimes \left( \hat{\mathbf{h}}_k^{\sf DL}(\hat{\mathbf{h}}_k^{\sf DL})^{\sf H} + \mathbf{\Phi}_k^{\sf}\right),
\end{align}
 where \(\mathbf{1}_k\) is an indicator vector whose \(k\)th element is \(1\) and the other elements are \(0\). The first-order optimality conditions is established in \cite{choi2019joint}. To completeness, we resate the first-order optimality condition as follows:
 \begin{thm}\label{thm: 1storder_opt}
    Let $\gamma(\mathbf{f}) = \prod_{k=1}^K \frac{\mathbf{f}^{\sf H}\mathbf{A}_k\mathbf{f}}{\mathbf{f}^{\sf H}\mathbf{B}_k\mathbf{f}}$. A stationary point $\mathbf{f} \in \mathbb{C}^{NK\times 1}$ for problem $\mathscr{P}_2$ is an eigenvector of the following functional generalized eigenvalue problem:
    \begin{align}
        \bar{\mathbf{A}}(\mathbf{f}) = \gamma(\mathbf{f})  \bar{\mathbf{B}}(\mathbf{f}),
    \end{align}
    where
    \begin{align}
        \bar{\mathbf{A}}(\mathbf{f}) = \sum_{k=1}^K\left( \prod_{k\ne i} \mathbf{f}^{\sf H}\mathbf{A}_k\mathbf{f}     \right)\mathbf{A}_i,
    \end{align}
    and 
    \begin{align}
        \bar{\mathbf{B}}(\mathbf{f}) = \sum_{k=1}^K\left( \prod_{k\ne i} \mathbf{f}^{\sf H}\mathbf{B}_k\mathbf{f}     \right)\mathbf{B}_i,
    \end{align}
\end{thm}
\begin{IEEEproof}
    See \cite{choi2019joint}.
\end{IEEEproof}
\vspace{0.2cm}

In general, finding the principal eigenvector \({\bf f} \in \mathbb{C}^{NK\times 1}\) that satisfies the first-order optimality condition \(\bar{\mathbf{A}}(\mathbf{f}) = \gamma(\mathbf{f}) \bar{\mathbf{B}}(\mathbf{f})\) is challenging because the matrices \(\bar{\mathbf{A}}(\mathbf{f})\) and \(\bar{\mathbf{B}}(\mathbf{f})\) are also functions of \({\bf f}\). The GPIP algorithm, developed in \cite{choi2019joint, han2023fdd}, is a computationally efficient method that identifies the precoding solution ensuring the first-order optimality condition. The overall algorithm is summarized in Algorithm \ref{Alg: GPIP}. In each iteration, the algorithm computes the matrices $\bar{\mathbf{A}} (\mathbf{f}^{(t-1)})$ and $\bar{\mathbf{B}} (\mathbf{f}^{(t-1)})$ based on $\mathbf{f}^{(t-1)}$. Subsequently, $\mathbf{f}^{(t)}$ is updated using the previously determined $\bar{\mathbf{A}} (\mathbf{f}^{(t-1)})$ and $\bar{\mathbf{B}} (\mathbf{f}^{(t-1)})$. This iterative process continues until the stopping criterion $|\gamma(\mathbf{f}^{(t-1)}) -\gamma(\mathbf{f}^{(t)})|/\gamma(\mathbf{f}^{(t-1)}) \ge \epsilon $ is satisfied. We refer to more details about the convergence properties of the GPIP algorithm as in \cite{han2023fdd, choi2019joint}. The complexity of GPIP is order of $\mathcal{O}(JN^2K)$, where $J$ is the number of iterations, which is typically less than 5 for the massive MIMO setting $N=K=64$ \cite{choi2019joint,GPIP2}.

\begin{algorithm}[t]
    \caption{Proposed Robust DL Precoding Algorithm (GPIP) \cite{han2023fdd}}
    Initialization: $t=1,~ \mathbf{f}^{(0)} = \mathbf{0},~ \mathbf{f}^{(0)} = \mathbf{f}^{\sf ZF}$\;
    \While{$t \ne 0$}{
    $t = t+1$\;
    \eIf{$|\gamma(\mathbf{f}^{(t-1)}) -\gamma(\mathbf{f}^{(t)})|/\gamma(\mathbf{f}^{(t-1)}) \ge \epsilon $  }
    {$\mathbf{f}^{(t)} \leftarrow \left(\bar{\mathbf{B}} (\mathbf{f}^{(t-1)})  \right)^{-1}   \bar{\mathbf{A}} (\mathbf{f}^{(t-1)}) \mathbf{f}^{(t-1)} $\;
    $\mathbf{f}^{(t)} \leftarrow \frac{\mathbf{f}^{(t)}}{\Vert\mathbf{f}^{(t)}\Vert_2}$\;    
    }
    {$\mathbf{f}^\star \leftarrow  \mathbf{f}^{(t)}$ \;}    
    }
    \label{Alg: GPIP}
\end{algorithm}

\section{Simulation Results}\label{Sec: Simulation Results}
In this section, we provide system-level simulation results to compare the ergodic DL sum-spectral efficiency across various precoding methods and CSI assumptions at the BS. Simulation parameters are summarized in Table \ref{tab: Params}.

\begin {table}[t]
\footnotesize
\caption {Simulation Parameters} \vspace{-0.3cm}\label{tab: Params} 
  	 \begin{center}
  \begin{tabular}{ l  c }
    \hline\hline
    \multicolumn{2}{c}{Parameters}   \\ \hline
        BS topology & Single hexagonal cell with ISD 500m \\
        User distribution & Uniform   \\ 
        UL carrier frequency & 10 GHz \\
        DL carrier frequency & 12 GHz \\
        The number of users $K$ & $16$\\
        Noise power & -113dBm\\
        Path-loss model & Standard model at TR 38.901\\\hline
  \end{tabular}
\end{center}\vspace{-0.3cm}
\end {table}

  

\subsection{Effects of CSIT knowledge and Precoding Algorithm}

\begin{figure}[t]
	\centering
    \includegraphics[width=\linewidth]{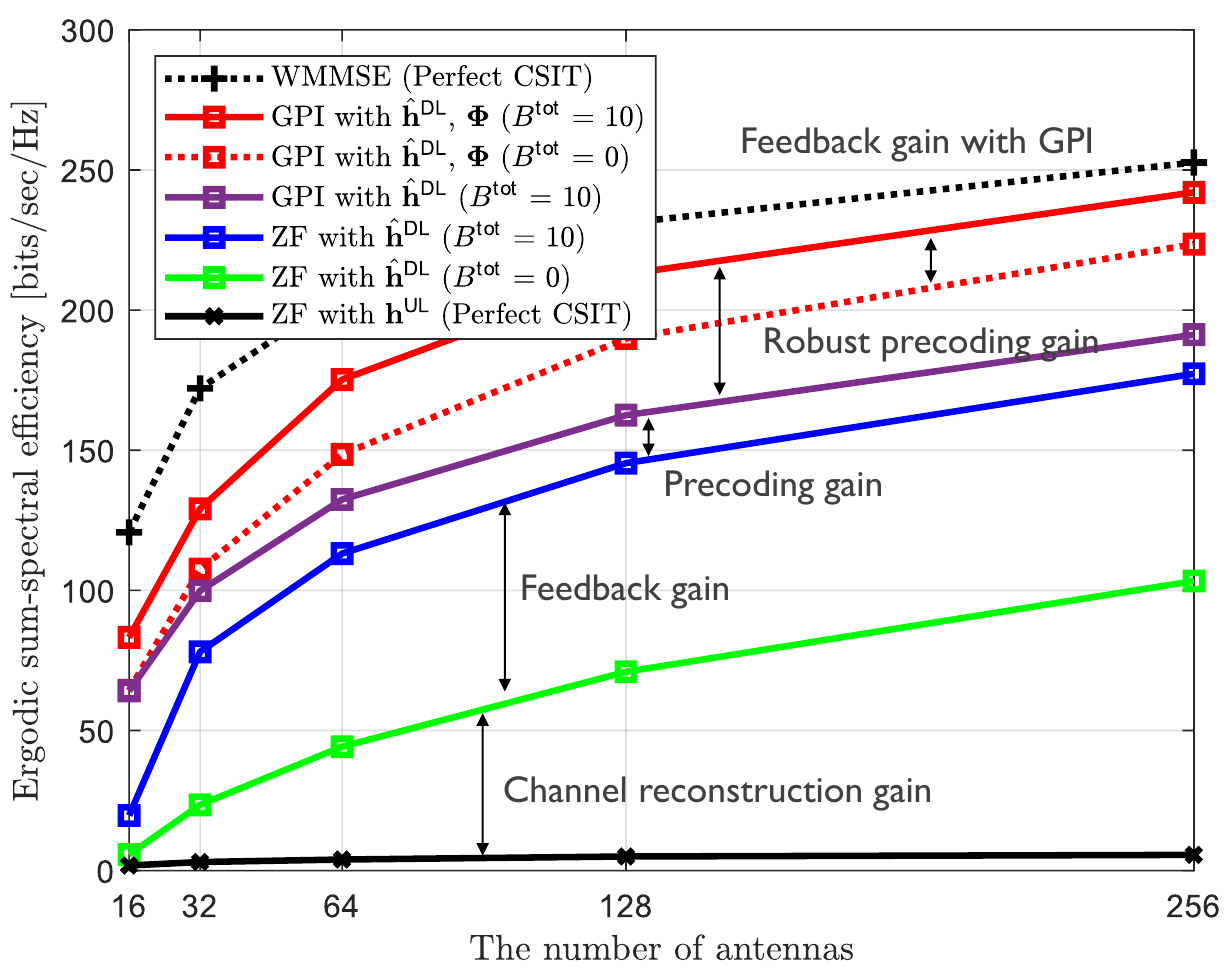}
  \caption{The ergodic sum-spectral efficiency with $(L_k,K) = (3,16)$ according to the CSIT knowledge and precoding algorithms.}\label{fig: results_Main1_CSIandPre}
\end{figure}

Fig. \ref{fig: results_Main1_CSIandPre} compares the ergodic DL sum-spectral efficiency of various precoding algorithms based on different levels of CSI knowledge at the BS, with settings $(K,L_k)=(16,3)$. The CSI knowledge levels at the BS are categorized as follows:
 \begin{itemize}
     \item Perfect UL channel knowledge:  \({\bf h}_k^{\sf UL} =   \sum_{\ell=1}^{L_k} g_{k,\ell}^{\sf UL}{\bf a}(\theta_{k,\ell}, \lambda^{\sf UL})\);
     \item DL channel reconstruction using the path gains and AoAs: \({\bf \hat h}_k^{\sf DL} = \sum_{\ell=1}^{L_k} \beta_{k,\ell}{\bf a}(\theta_{k,\ell}, \lambda^{\sf DL})\);
     \item DL channel reconstruction using the path gains, AoAs, and $B^{\rm tot}$-bit phase feedback:  \({\bf \hat h}_k^{\sf DL} = \sum_{\ell=1}^{L_k}\eta_{k,\ell} \beta_{k,\ell} e^{jq_{k,\ell}^{\sf DL}}{\bf a}(\theta_{k,\ell}, \lambda^{\sf DL})\);
    \item DL channel reconstruction using the path gains, AoAs,  $B^{\rm tot}$-bit phase feedback, and the error covariance matrix:  \({\bf \hat h}_k^{\sf DL} = \sum_{\ell=1}^{L_k} \eta_{k,\ell}\beta_{k,\ell} e^{jq_{k,\ell}^{\sf DL}}{\bf a}(\theta_{k,\ell}, \lambda^{\sf DL})\) and ${\bf \Phi}_{k,\ell}$.
\end{itemize}
 
We first examine the DL sum spectral efficiency based on different levels of CSI knowledge when ZF precoding is applied. It is evident that the sum-spectral efficiency increases as the BS gains more accurate DL CSI knowledge. Notably, even a few bits of CSI feedback per DL channel path can significantly enhance performance.

Next, we evaluate the DL performance when GPIP-based robust precoding is applied. As observed in previous studies \cite{choi2019joint,han2023fdd}, utilizing the error covariance matrix in conjunction with the reconstructed DL channel effectively improves sum-spectral efficiency. This improvement occurs because the joint use of these elements enables a more accurate estimation of \({\bf h}_k^{\sf DL}({\bf h}_k^{\sf DL})^{\sf H}\), as demonstrated in Lemma \ref{lem3}. Therefore, accurately identifying the error covariance matrix is crucial for enhancing DL channel reconstruction. We also note that the proposed DL channel reconstruction combined with the robust precoding method can achieve performance levels comparable to the upper bound attained by WMMSE precoding with perfect DL CSI knowledge.

 


\subsection{Effect of the Codebook Design}
\begin{figure}[t]
	\centering
    \includegraphics[width=\linewidth]{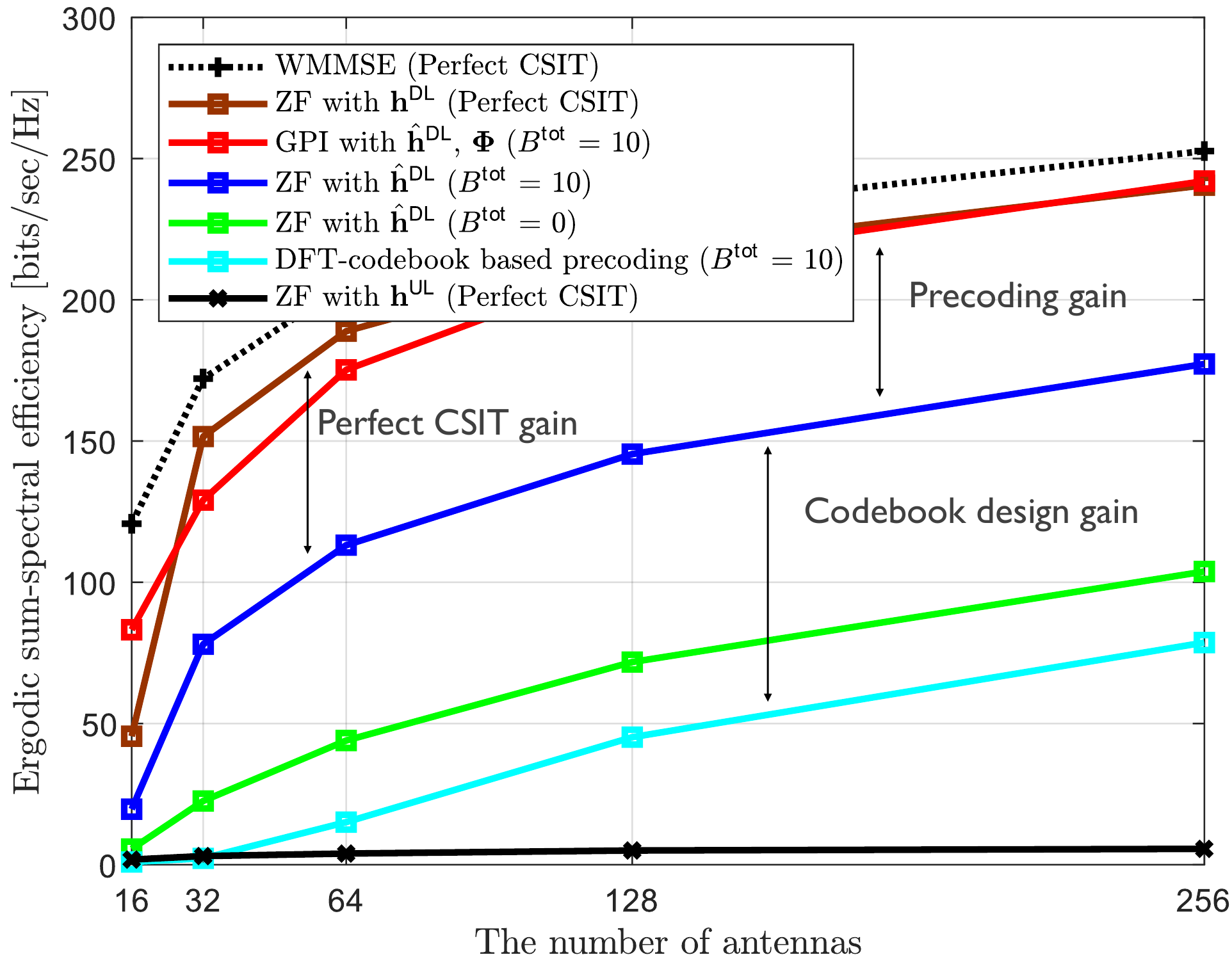}
  \caption{The ergodic sum-spectral efficiency with $(L_k,K) = (3,16)$ according to the CSIT knowledge and codebook design.}\label{fig: results_Main6_RVQ}
\end{figure}

Fig. \ref{fig: results_Main6_RVQ} compares the ergodic DL sum-spectral efficiency of the discrete Fourier transform (DFT) codebook-based CSI feedback method to the proposed channel path phase feedback method, with settings $(K,L_k)=(16,3)$. DFT codebook-based CSI feedback is a conventional CSI feedback method in which a vector from the DFT matrix that has the largest inner product with the DL channel is selected and provided as feedback to the BS \cite{DFT1}. Simulation results show that utilizing the information of AoA and path gain, which are reciprocal in FDD systems, for DL channel reconstruction brings ergodic sum-spectral efficiency gains related to codebook design.

\subsection{Effect of the Number of Feedback Bits}
\begin{figure}[t]
	\centering
    \includegraphics[width=\linewidth]{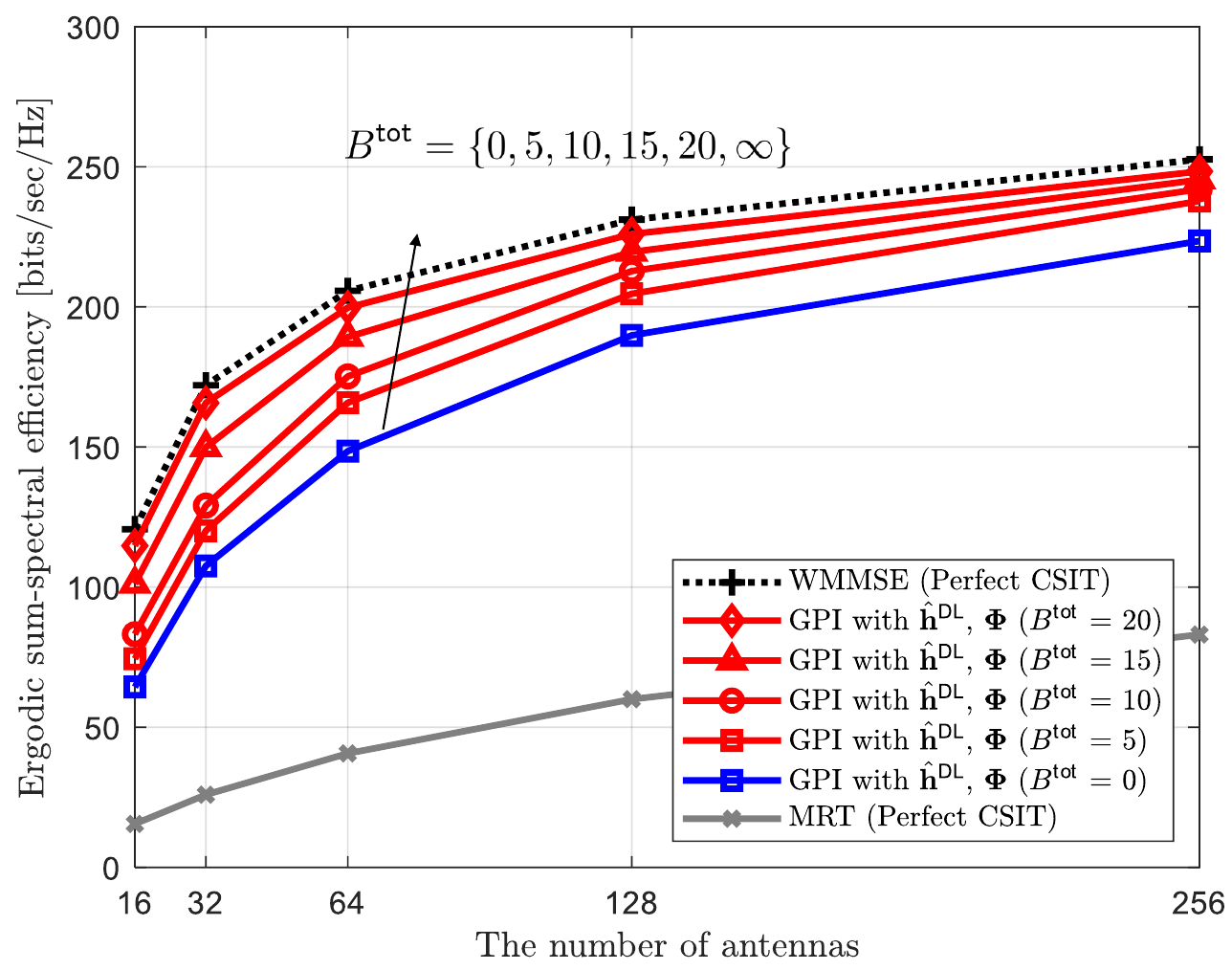}
  \caption{The ergodic sum-spectral efficiency $(L_k,K) = (3,16)$ as increasing the number of feedback bits per user $B^{\sf tot}$.}\label{fig: results_Main2_Bits}
\end{figure}

Fig. \ref{fig: results_Main2_Bits} illustrates the DL ergodic sum-spectral efficiency as the total number of feedback bits, $B^{\sf tot}$, increases. According to Theorem \ref{thm:MSE} and Lemma \ref{lem1}, we have demonstrated that the DL channel reconstruction MSE value decreases with an increase in $B^{\sf tot}$, indicating that the BS gains more accurate DL channel knowledge. As expected, the DL sum-spectral efficiency also improves with increasing $B^{\sf tot}$. It is noteworthy that when $(K, L_k) = (16, 3)$, providing just 20 CSI feedback bits per user can closely achieve the upper bound performance achieved by WMMSE precoding with perfect DL CSI knowledge, regardless of the number of antennas.

\subsection{Effects of the Geometric Parameter Estimation Errors}
\begin{figure}[t]
	\centering
    \includegraphics[width=\linewidth]{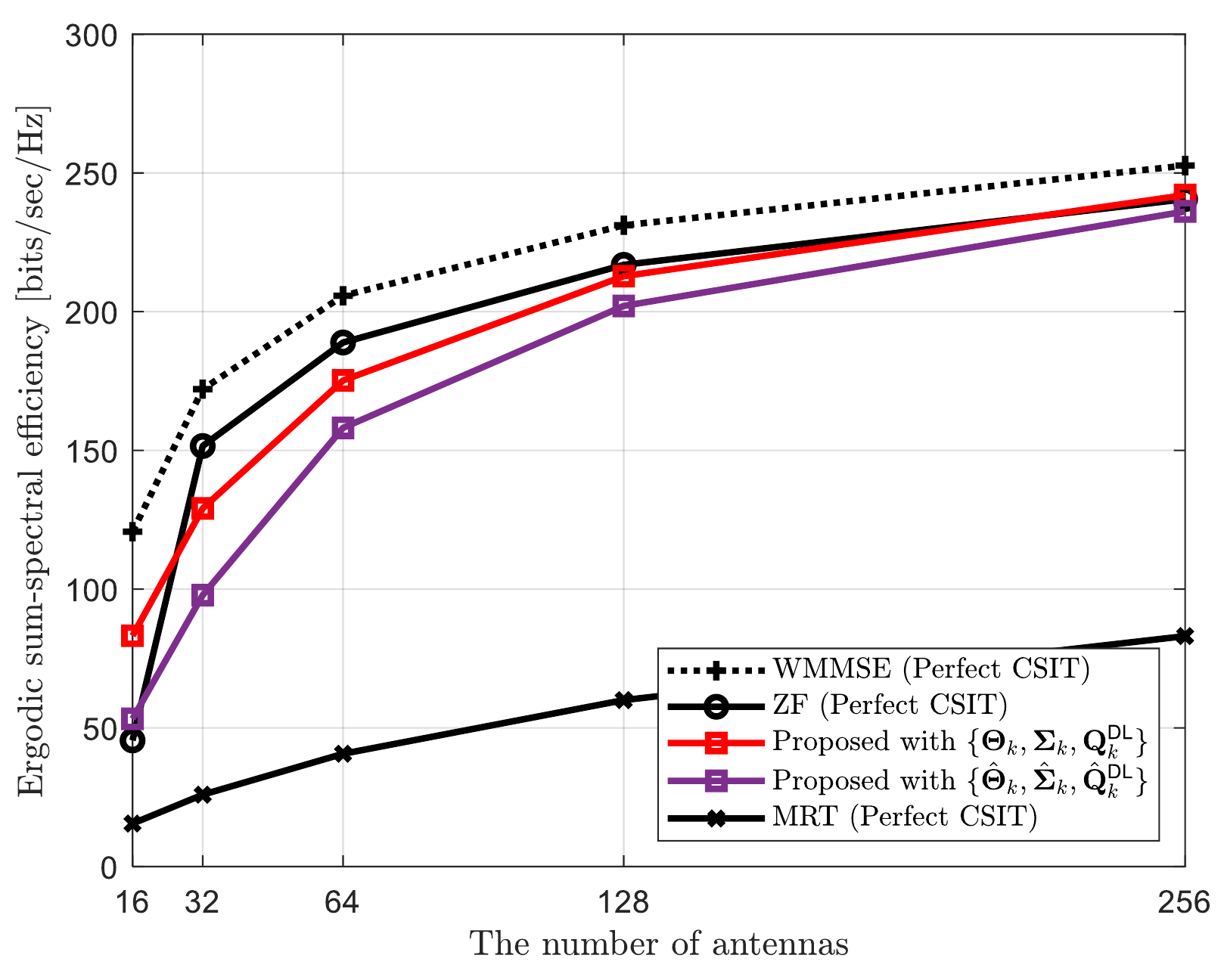}
  \caption{The ergodic sum-spectral efficiency with $(L_k,K) = (3,16)$ in relation to channel parameter estimation errors.}\label{fig: results_Main4_ParamEstErr}
\end{figure}

 Fig. \ref{fig: results_Main4_ParamEstErr} shows the DL ergodic sum-spectral efficiency as the number of antennas increases. The proposed algorithm is evaluated under two conditions: One with perfect channel parameter knowledge, represented as \({\bf \hat h}_k^{\sf DL} = \sum_{\ell=1}^{L_k} \eta_{k,\ell} \beta_{k,\ell} e^{jq_{k,\ell}^{\sf DL}}{\bf a}(\theta_{k,\ell}, \lambda^{\sf DL})\) and ${\bf \Phi}_{k,\ell}$, and another with imperfect channel parameter knowledge, where \({\bf \hat h}_k^{\sf DL} = \sum_{\ell=1}^{L_k} \eta_{k,\ell}{\hat \beta}_{k,\ell} e^{jq_{k,\ell}^{\sf DL}}{\bf a}({\hat \theta}_{k,\ell}, \lambda^{\sf DL})\) and ${\bf \Phi}_{k,\ell}$. The channel parameters in both UL and DL are estimated using the spatial smoothing MUSIC technique \cite{Spatial} on noisy received signals from the UL and DL pilots under the system configuration in Table \ref{tab: Params}. As anticipated, the sum-spectral efficiency is slightly compromised by the effects of parameter estimation. However, the performance degradation is relatively minor. Notably, when the BS is equipped with a sufficient number of antennas, the accuracy of parameter estimation improves, resulting in negligible performance degradation.

\begin{figure}[t]
	\centering
    \includegraphics[width=\linewidth]{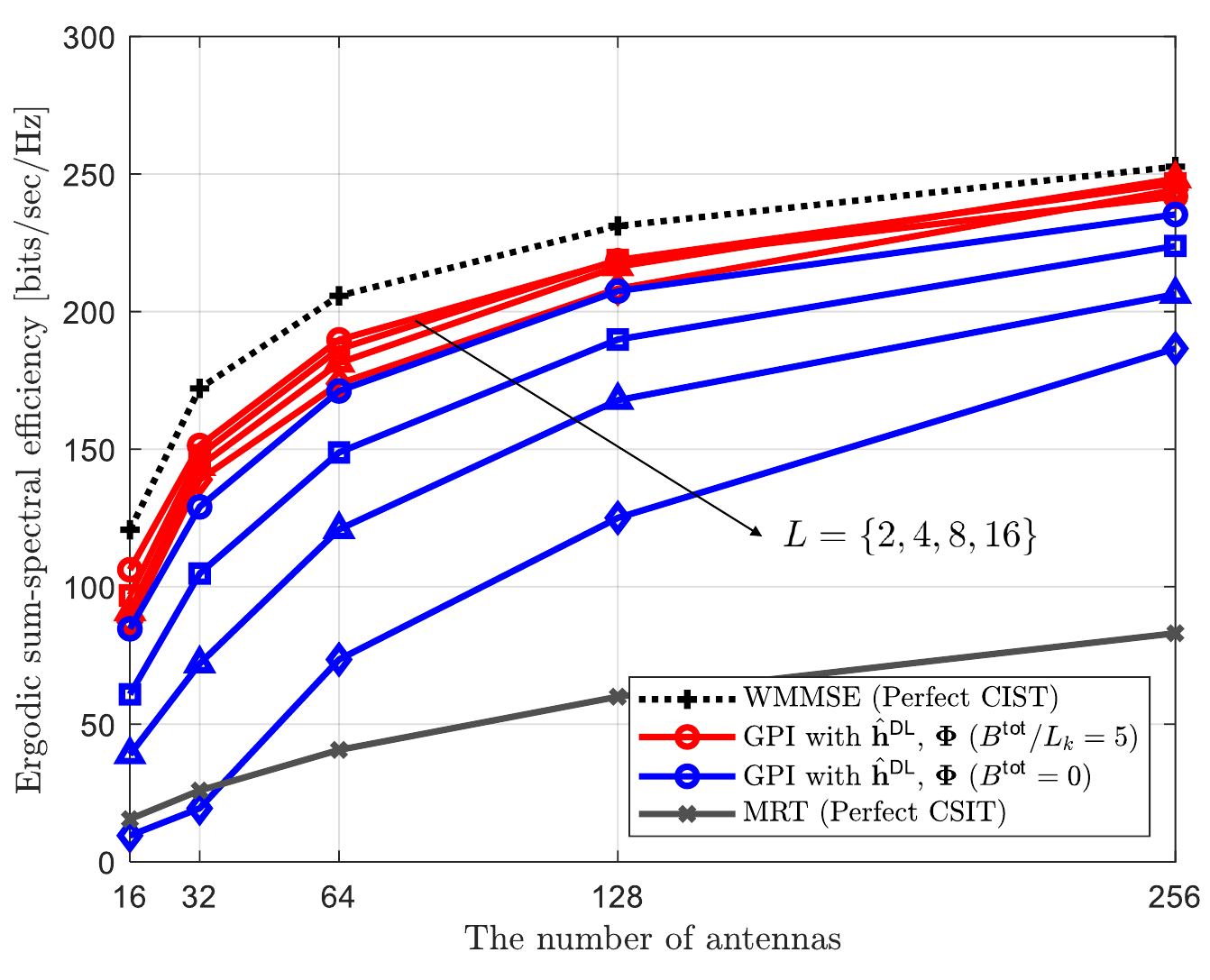}
  \caption{The ergodic sum-spectral efficiency as increasing the number of channel paths $L_k$. The number of feedback bits used per channel path is set to $B^{\sf tot}/L_k = 5$. The number of users is set to $K = 16$.}\label{fig: results_Main3_Paths}
\end{figure}

\subsection{Effect of the Number of Channel Paths}
Fig. \ref{fig: results_Main3_Paths} illustrates the DL ergodic sum-spectral efficiency as the number of channel paths per user increases. As the number of channel paths grows, DL channel reconstruction performance deteriorates. This decline is primarily due to the increased error in estimating the geometric parameters, as a greater number of parameters leads to decreased accuracy in channel reconstruction. However, in the proposed DL data transmission structure, the degradation in ergodic spectral efficiency is significantly less when using feedback compared to not using any. This is because fixing the number of feedback bits per multipath channel path allows for an accurate depiction of the complex channel environments. Consequently, if a certain level of feedback is used per channel path is guaranteed, a robust DL data transmission framework against channel complexity can be established.

\section{Conclusion}\label{Sec: Conclusion}
We introduced a novel DL data transmission framework in FDD massive MIMO systems that can yield significant gains in sum spectral efficiency. Our approach involves utilizing information on channel parameters that maintain reciprocity between UL and DL channels, and feedback for the channel parameters that do not maintain reciprocity to reconstruct the optimal DL channel from an MSE perspective. The MSE of the reconstructed channel is ultimately influenced by the number of bits used per path in the feedback by the users. Consequently, we have defined and addressed the optimal bit allocation problem for each path within a given bit budget per user. Furthermore, we proposed a robust DL precoding technique that maximizes sum spectral efficiency, leveraging the reconstructed DL channel and its corresponding error covariance matrix. Our key finding demonstrates that FDD massive MIMO system can achieve significant gains with minimal feedback on the phase of channel paths.  

One possible extension of this work involves exploring vector quantization techniques for limited phase information feedback, which could enhance the MSE performance of DL channel reconstruction. Another potential research direction is to expand the proposed algorithm by incorporating data-driven machine learning methods.

\appendix


    



\subsection{Proof for Theorem \ref{thm: MSE-OptDLCh}}\label{Prf: MSE-OptDLCh}
The optimal DL channel estimator with a knowledge of $\{\mathbf{h}_k^{\sf UL}, \mathbf{\Theta}_k, \mathbf{\Sigma}_k, \mathbf{Q}_k^{\sf DL}\}$ can be calculated by the following conditional expectation:
\begin{align}
    \hat{\mathbf{h}}_k^{\sf DL, MMSE} & = \mathbb{E} \left[\mathbf{h}_k^{\sf DL} \ \vert\  \mathbf{h}_k^{\sf UL}, \mathbf{\Theta}_k, \mathbf{\Sigma}_k, \mathbf{Q}_k^{\sf DL} \right] \nonumber \\
    &= \mathbf{A}_k^{\sf DL} \mathbb{E} \left[ \mathbf{g}_k^{\sf DL} \ \vert\  \mathbf{h}_k^{\sf UL}, \mathbf{\Theta}_k, \mathbf{\Sigma}_k, \mathbf{Q}_k^{\sf DL}   \right], \label{eq: CH-MMSE}
\end{align}
where the matrix of array response vectors $\mathbf{A}_k^{\sf DL} = [\mathbf{a}(\theta_{k,1}, \lambda^{\sf DL}) , \cdots, \mathbf{a}(\theta_{k,L_k}, \lambda^{\sf DL}) ] \in \mathbb{C}^{N\times L_k}$, the vector of channel path coefficients $\mathbf{g}_k^{\sf DL} = [ g_{k,1}^{\sf DL}, \cdots, g_{k,L_k}^{\sf DL} ]^\top \in \mathbb{C}^{L_k\times 1 }$, the matrix of AoAs $\mathbf{\Theta}_k = {\sf diag} \left(\left[\theta_{k,1}, \cdots, \theta_{k,L_k} \right]\right) \in \mathbb{R}^{L_k\times L_k }$, the matrix of channel path gains $\mathbf{\Sigma}_k = {\sf diag}\left(\left[ \beta_{k,1},\cdots, \beta_{k,L_k} \right]\right) \in \mathbb{R}^{L_k\times L_k}$, and the matrix of quantized feedback phase $\mathbf{Q}_k^{\sf DL} = {\sf diag}([ e^{jq_{k,1}^{\sf DL}},\cdots, e^{jq_{k,L_k}^{\sf DL}} ]) \in \mathbb{C}^{L_k\times L_k}$.

For easy notation, we denote the path attenuation vector $\pmb{\beta}_k = \left[\beta_{k,1},\cdots,\beta_{k,L_k} \right]^\top$, the quantized phase vector $e^{j \mathbf{q}_k^{\sf DL}} = \left[e^{jq_{k,1}^{\sf DL}}, \cdots, e^{jq_{k,L_k}^{\sf DL}} \right]^\top $, and the feedback error vector  $e^{j\pmb{\delta}_k} = \left[e^{j\delta_{k,1}}, \cdots, e^{j\delta_{k,L_k}} \right]^\top $. As a result, the DL channel coefficients $\mathbf{g}_k^{\sf DL} = \pmb{\beta}_k \odot e^{j \mathbf{q}_k^{\sf DL}} \odot e^{j\pmb{\delta}_k} $. Using these notations, the expectation in (\ref{eq: CH-MMSE}) is obtained over the feedback error term:
\begin{align}
    &\mathbb{E} \left[  \mathbf{g}_k^{\sf DL} \ \vert\  \mathbf{h}_k^{\sf UL}, \mathbf{\Theta}_k, \mathbf{\Sigma}_k, \mathbf{Q}_k^{\sf DL}   \right] \nonumber\\
    & \ =  \pmb{\beta}_k  \odot e^{j \mathbf{q}_k^{\sf DL}} \odot \int e^{j\pmb{\delta}_k} f\left(\pmb{\delta}_k| \mathbf{h}_k^{\sf UL}, \mathbf{\Theta}_k, \mathbf{\Sigma}_k, \mathbf{Q}_k^{\sf DL}    \right) d \pmb{\delta}_k \nonumber \\
    &\ = \pmb{\beta}_k \odot e^{j \mathbf{q}_k^{\sf DL}} \odot \pmb{\eta}_k. \label{eq: eta}
\end{align}
The expected feedback error in (\ref{eq: eta}) is calculated with the feedback bits of the $\ell$th path as
\begin{align}
     \eta_{k,\ell} &=  \int_{-\frac{2\pi}{2^{B_{k,\ell}+1}}}^{\frac{2\pi}{2^{B_{k,\ell}+1}}} e^{j\delta_{k,\ell}}  \frac{2^{B_{k,\ell}}}{2\pi} d \delta_{k,\ell} \nonumber \\
     &= j\frac{2^{B_{k,\ell}}}{2\pi} e^{-j \frac{2\pi}{2^{B_{k,\ell}+1}}} -  j\frac{2^{B_{k,\ell}}}{2\pi} e^{j \frac{2\pi}{2^{B_{k,\ell}+1}}} \nonumber \\ 
     &= \frac{2^{B_{k,\ell}}}{\pi} \sin \left( \frac{\pi}{2^{B_{k,\ell}}}\right). \label{eq: eta_detail}
\end{align}
Consequently, invoking (\ref{eq: eta}) into (\ref{eq: CH-MMSE}), we obtain the MSE-optimal estimator as
\begin{align}
    \hat{\mathbf{h}}_k^{\sf DL, MMSE} &= \mathbf{A}_k^{\sf DL} \mathbf{\Sigma}_k \mathbf{Q}_k^{\sf DL} \pmb{\eta}_k \nonumber \\
    &= \sum_{\ell=1}^{L_k} \eta_{k,\ell}\beta_{k,\ell}e^{jq_{k,\ell}^{\sf DL}} \mathbf{a}\left(\theta_{k,\ell}, \lambda^{\sf DL}\right),
\end{align}
which completes the proof of Theorem \ref{thm: MSE-OptDLCh}.

\subsection{Proof for Theorem \ref{thm:MSE}}\label{Prf: MSE}
When the DL channel parameters $\mathbf{h}_k^{\sf UL}, \mathbf{\Theta}_k, \mathbf{\Sigma}_k, \mathbf{Q}_k^{\sf DL}$ given, the conditional MSE is computed as
\begin{align}
    {\sf MSE}_k &= \mathbb{E}\left[ \left. \left\Vert\mathbf{h}_k^{\sf DL} - \hat{\mathbf{h}}_k^{\sf DL, MMSE} \right\Vert_2^2 \right|\mathbf{h}_k^{\sf UL}, \mathbf{\Theta}_k, \mathbf{\Sigma}_k, \mathbf{Q}_k^{\sf DL} \right] \nonumber \\
    &= \mathbb{E}\left[ \left. \sum_{\ell=1}^{L_k} \left\Vert \mathbf{h}_{k,\ell}^{\sf DL} - \hat{\mathbf{h}}_{k,\ell}^{\sf DL, MMSE} \right\Vert_2^2 \right|\mathbf{h}_k^{\sf UL}, \mathbf{\Theta}_k, \mathbf{\Sigma}_k, \mathbf{Q}_k^{\sf DL}\right] \nonumber\\
    &=\mathbb{E}\left[\sum_{\ell=1}^{L_k} \left\Vert \beta_{k,\ell}\left( e^{j\angle g_{k,\ell}^{\sf DL}} - \eta_{k,\ell}e^{jq_{k,\ell}^{\sf DL}} \right) \mathbf{a}\left(\theta_{k,\ell}^{\sf DL}, \lambda^{\sf DL}\right) \right\Vert_2^2  \right] \nonumber \\
    &=  \sum_{\ell=1}^{L_k} N \beta_{k,\ell}^2 \mathbb{E} \left[ \left| e^{j\angle g_{k,\ell}^{\sf DL}} - \eta_{k,\ell}e^{jq_{k,\ell}^{\sf DL}} \right|^2 \right],\label{eq: Apx C_MSE}
\end{align}
where $\mathbf{h}_{k,\ell}^{\sf DL} = \beta_{k,\ell}e^{j\angle g_{k,\ell}^{\sf DL}}\mathbf{a}\left(\theta_{k,\ell}^{\sf DL}, \lambda^{\sf DL}\right)$ and $\hat{\mathbf{h}}_{k,\ell}^{\sf DL, MMSE} =\eta_{k,\ell}\beta_{k,\ell}e^{jq_{k,\ell}^{\sf DL}}\mathbf{a}\left(\theta_{k,\ell}^{\sf DL}, \lambda^{\sf DL}\right) $ are the $\ell$th channel path for the true and the reconstructed DL channels. The second equality holds because $\mathbf{h}_{k,\ell}^{\sf DL}$ and $\mathbf{h}_{k,\ell'}^{\sf DL}$ are assumed to be statistically uncorrelated $\mathbb{E}\left[ \mathbf{h}_{k,\ell}^{\sf DL}\left( \mathbf{h}_{k,\ell'}^{\sf DL}\right)^{\top}\right]={\bf 0}$ for $\ell\neq \ell'$. To this end, we need to compute he expectation in (\ref{eq: Apx C_MSE}) as
\begin{align}
  &  \mathbb{E} \left[ \left| e^{j\angle g_{k,\ell}^{\sf DL}} - \eta_{k,\ell}e^{jq_{k,\ell}^{\sf DL}} \right|^2 \right]\nonumber \\
    &=    \mathbb{E} \left[  \left( 1 -\eta_{k,\ell}\left( e^{j\delta_{k,\ell}}+e^{-j\delta_{k,\ell}}   \right) + \eta_{k,\ell}^2 \right) \right] \nonumber \\ 
    &=    1 - \eta_{k,\ell}^2   \nonumber \\
    &=    1 - \left(\frac{2^{B_{k,\ell}}}{\pi} \sin \left(  \frac{\pi}{2^{B_{k,\ell}}}   \right) \right)^2. \label{eq:expect}
\end{align}
Invoking \eqref{eq:expect} into \eqref{eq: Apx C_MSE}, we get the MSE expression in Theorem \ref{thm:MSE}, which completes the proof.

\subsection{Proof for Lemma \ref{lem1}}\label{Prf: convex_dec}
If the function ${\tilde f}^{\sf MSE}_{k,\ell}$ is strictly decreasing, the following inequality always holds:
\begin{align}
    {\tilde f}^{\sf MSE}_{k,\ell}(x) > {\tilde f}^{\sf MSE}_{k,\ell}(y) ~~ \forall x<y. \label{eq: Proof D_1}
\end{align}
However, the domain of \({\tilde f}^{\sf MSE}_{k,\ell}\) is \(\mathbb{R}_{\ge 0}\). Additionally, within the domain corresponding to \(\mathbb{Z}_{\ge 0}\), \({\tilde f}^{\sf MSE}_{k,\ell}\) yields the same values as \(f^{\sf MSE}_{k,\ell}\), and for other points, it is linearly interpolated. Consequently, the inequality \eqref{eq: Proof D_1} consistently holds when the following inequality is satisfied:
\begin{align}
    f^{\sf MSE}_{k,\ell}(x) > f^{\sf MSE}_{k,\ell}(x+1) ~~ \forall x\in \mathbb{Z}_{\ge 0}. \label{eq: Proof D_2}
\end{align}
Then, the difference in function values between two adjacent integers is defined as follows:
\begin{align}
    \Delta^{\sf MSE}(x) &= f^{\sf MSE}_{k,\ell}(x) - f^{\sf MSE}_{k,\ell}(x+1) \nonumber \\
    &= \frac{4^x}{\pi^2}\left(\sin^2\left(\frac{\pi}{2^x}\right) - 4 \sin^2\left(\frac{\pi}{2^{x+1}}\right)       \right)\nonumber \\
    &= \frac{4^x}{\pi^2}\left(1 - \cos^2\left( \frac{\pi}{2^x} \right) -2+\cos\left(\frac{\pi}{2^x}\right)\right) \nonumber \\
    &= -\frac{4^x}{\pi^2}\left(1 - \cos\left( \frac{\pi}{2^x} \right)\right)^2 .\label{eq: Proof D_3}
\end{align}
 As shown in \eqref{eq: Proof D_3}, \(\Delta^{\sf MSE}(x)\) is always less than zero, indicating that \({\tilde f}^{\sf MSE}_{k,\ell}\) is a strictly decreasing function.
 
If ${\tilde f}^{\sf MSE}_{k,\ell}$ is a convex function, the following inequality always holds for all $x$ and $y$:
\begin{align}
    {\tilde f}^{\sf MSE}_{k,\ell}\left( ax + (1-a)y\right) \le a {\tilde f}^{\sf MSE}_{k,\ell}(x) + (1-a){\tilde f}^{\sf MSE}_{k,\ell}(y), \label{eq: Proof D_4}
\end{align}
where $a \in [0,1]$. However, for the same reasons demonstrated above when establishing that the function is strictly decreasing, the inequality \eqref{eq: Proof D_4} is always valid when the following condition is met:
\begin{align}
     2\tilde{f}^{\sf MSE}_{k,\ell}(x+1) \le \tilde{f}^{\sf MSE}_{k,\ell}(x) + \tilde{f}^{\sf MSE}_{k,\ell}(x+2) ~~ \forall x\in \mathbb{Z}_{\ge 0}. \label{eq: Proof D_5}
\end{align}
The inequality in \eqref{eq: Proof D_5} can be rewritten as
\begin{align}
    \sin^2\left(\frac{\pi^2}{2^{x}}\right) +16\sin^2\left(\frac{\pi^2}{2^{x+2}}\right) - 4\sin^2\left(\frac{\pi^2}{2^{x+1}}\right) \le 0.
\end{align}
Let $\frac{\pi^2}{2^{x+1}} = A$, then the above inequality becomes
\begin{align}
    0 &\ge \sin^2(2A) + 16 \sin^2(\frac{A}{2}) - 8\sin^2(A) \nonumber \\
    &= 8\sin^2 A -\sin^2(2A) - 16 \sin^2(A/2)  \nonumber \\
    &=8(1-\cos^2 A) -  4(1-\cos^2 A)\cos^2 A - 8(1-\cos A)  \nonumber\\
    &=4\cos A(1-\cos A)(2-\cos A - \cos^2 A),\label{eq: Proof D_6}
\end{align}
Because $-1\le\cos(A)\le 1$, the inequality \eqref{eq: Proof D_6} always holds. Consequently, ${\tilde f}^{\sf MSE}_{k,\ell}$ is a convex function. This completes the proof.

\subsection{Proof for Theorem \ref{thm: bit_allo_optimality}}\label{Prf: bit_allo_optimality}
To demonstrate the optimality of the bit allocation determined by Algorithm \ref{Alg: Opt Bit Allo}, we provide a proof using the discrete optimization technique introduced in \cite{fox1966discrete}. For clarity and ease of mathematical expression, we define the following optimization problem, which is equivalent to problem $\mathscr{P}1^\prime$:
\begin{align}
    & ~\underset{x_1, \cdots, x_n}{\rm max} \sum_{j=1}^n  \phi_j(x_j) \\
    &~\text{subject to}~ \sum_{j=1}^n x_j \le M \\
    &~\text{subject to}~ x_j \in \mathbb{Z}_{\ge 0}. \label{eq: Proof E_1}    
\end{align}
Define \(\phi(x) = \sum_{j=1}^n  \phi_j(x_j)\) and \(C(x) = \sum_{j=1}^n x_j\). In this optimization problem, the allocation of \(x = [x_1, \cdots, x_n]\) must satisfy the following conditions:\begin{align}
    \phi(y) > \phi(x) \Rightarrow  C(y) > C(x) \nonumber \\ 
    \phi(y) = \phi(x) \Rightarrow  C(y) \ge C(x). 
\end{align}
 An allocation \(y\) is called undominated if it satisfies certain criteria. During the allocation process in Algorithm \ref{Alg: Opt Bit Allo}, it is evident that \(x^{(0)}\) is undominated. Set \(\lambda_t\) to the value:
 \begin{align}
    \lambda_t = \underset{j}{\rm max} \ \ \phi_j\left(x_j^{(t-1)}+1\right) - \phi_j\left(x_j^{(t-1)}\right). \label{eq: Proof E_2}
\end{align}
If $\phi_j(x_j)$ is a strictly increasing and concave function, by the definition of the algorithm, $\lambda_t \le \lambda_{t-1}$. Then, $x^{(1)} \in L(\lambda_1)$. $L(\lambda)$ is defined by the following procedures. Let $m_i^*(\lambda)$ be the smallest nonnegative integer $m$ satisfying $\phi_i(m+1) - \phi_i(m)< \lambda$, and $E_i(\lambda)$ be the set of nonnegative integers $m$ for which $\phi_i(m+1) - \phi_i(m)= \lambda$. Then $L_i(\lambda) = \{m_i^*(\lambda)\} \cup E_i(\lambda)$. Note that $L_i(\lambda)$ contains the global maxima of $\phi_i(x_i) - \lambda x_i $ over nonnegative integers if $\phi_i(x_i)$ is concave. Then $L(\lambda) = [l_1, \cdots, l_n]$ where $l_i \in L_i(\lambda)$ for all $i$. By the definition of $L(\lambda)$, it is clear that $x^{(1)} \in L(\lambda_1)$. Whether or not $i$ is an index corresponding to a maximum value in \eqref{eq: Proof E_2}, the $x_i$ obtained through the Algorithm \ref{Alg: Opt Bit Allo} satisfies the following conditions: $x_i^{(t-1)} \in L_i(\lambda_{t-1}) \Rightarrow x_i^{(t)} \in L_i(\lambda_{t})$. So, $x^{(t)} \in L(\lambda_t)$

Then, for any $\lambda >0$, if $\phi_i(x_i)$ is concave for all $i$ and $x \in L(\lambda)$, $x$ is undominated. This can be understood through the Lagrange multiplier method. If $\lambda\ge0$ and $x^* \in \mathbb{Z}_{\ge 0}$ maximize the Lagrangian $\phi(x) - \lambda g(x)$ on all $x \in \mathbb{Z}_{\ge 0}$, then $x^*$ maximizes $\phi(x)$ over all $x$ such that $g(x)\le g(x^*)$. So $x \in L(\lambda)$ is the global maximum of $\phi(x)$. Therefore, $x$ is undominated and Algorithm \ref{Alg: Opt Bit Allo} outputs the optimal bit allocation.

 Our objective function, \(\beta_{k,\ell}{\tilde f}^{\sf MSE}_{k,\ell}\), is both convex and strictly decreasing. However, given that our optimization problem is one of minimization, we can reformulate it as a maximization problem using \(-\beta_{k,\ell}{\tilde f}^{\sf MSE}_{k,\ell}\). This adaptation allows us to apply this proof effectively, thereby completing the proof.

\subsection{Proof for Lemma \ref{lem2}}\label{Prf: Error cov}
In a similar manner to the (\ref{eq: MMSE DL_main}), the denote the DL channel that $\mathbf{h}_k^{\sf DL} = \mathbf{A}_k^{\sf DL} \mathbf{\Sigma}_k e^{j\angle\mathbf{g}_{k}^{\sf DL}}$, where $e^{j\angle\mathbf{g}_{k}^{\sf DL}} = \left[e^{j\angle g_{k,1}^{\sf DL}}, \cdots, e^{j\angle g_{k,1}^{\sf DL}} \right]^\top$. Then, the error covariance matrix is
\begin{align}
    \mathbf{\Phi}_k^{\sf MMSE} &= \mathbb{E} \left[ \mathbf{e}_k  \mathbf{e}_k ^{\sf H} |\mathbf{h}_k^{\sf UL}, \mathbf{\Theta}_k, \mathbf{\Sigma}_k, \mathbf{Q}_k^{\sf DL}\right] \nonumber \\
    &=  \mathbb{E} \left[ \mathbf{A}_k^{\sf DL} \mathbf{\Sigma}_k \left(e^{j\angle\mathbf{g}_{k}^{\sf DL}} - \mathbf{Q}_k^{\sf DL} \pmb{\eta}_k  \right)  \right. \nonumber \\
    &\quad \left.\left(e^{j\angle\mathbf{g}_{k}^{\sf DL}} - \mathbf{Q}_k^{\sf DL} \pmb{\eta}_k \right)^{\sf H}  \mathbf{\Sigma}_k
    \left(\mathbf{A}_k^{\sf DL}\right)^{\sf H} \right] \nonumber \\
    &= \mathbf{A}_k^{\sf DL} \mathbf{\Sigma}_k  \mathbb{E}\left[ \left(e^{j\angle\mathbf{g}_{k}^{\sf DL}} - \mathbf{Q}_k^{\sf DL} \pmb{\eta}_k \right)\right. \nonumber \\ 
    &\quad \left. \left(e^{j\angle\mathbf{g}_{k}^{\sf DL}} - \mathbf{Q}_k^{\sf DL} \pmb{\eta}_k \right)^{\sf H}      \right] \mathbf{\Sigma}_k   \left(\mathbf{A}_k^{\sf DL}\right)^{\sf H} \nonumber \\
    &= \mathbf{A}_k^{\sf DL} \mathbf{\Sigma}_k  \left( \mathbf{I} - \mathbf{E}_k \right) \mathbf{\Sigma}_k  \left(\mathbf{A}_k^{\sf DL}\right)^{\sf H} \nonumber \\
    &= \sum_{\ell=1}^{L_k} \beta_{k,\ell}^2 (1-\eta_{k,\ell}^2) \mathbf{a}\left(\theta_{k,\ell}, \lambda^{\sf DL}\right)\mathbf{a}\left(\theta_{k,\ell}, \lambda^{\sf DL}\right)^{\sf H},
\end{align}
where $\mathbf{E}_k = {\sf diag}\left(\left[\eta_{k,1}^2, \cdots, \eta_{k,L_k}^2 \right] \right)$. This completes the proof.

\subsection{Proof for Theorem \ref{lem3}}\label{Prf: Outer apprx}
Utilizing the properties of the trace operation, we can derive the following equations:
\begin{align}
    &\Vert\mathbf{\Delta}_k\Vert_F^2  \nonumber \\
    &= \left\Vert \mathbf{h}_k^{\sf DL} (\mathbf{h}_k^{\sf DL})^{\sf H} -\left( \hat{\mathbf{h}}_k^{\sf DL, MMSE}(\hat{\mathbf{h}}_k^{\sf DL, MMSE})^{\sf H} + \mathbf{\Phi}_k^{\sf}\right)\right\Vert_F^2 \nonumber \\
    &= \bigg\Vert \mathbf{A}_k^{\sf DL} \mathbf{\Sigma}_k  \bigg(e^{j\angle\mathbf{g}_{k}^{\sf DL}}\left(e^{j\angle\mathbf{g}_{k}^{\sf DL}}\right)^{\sf H}  - \mathbf{Q}_k^{\sf DL} \pmb{\eta}_k\left(\mathbf{Q}_k^{\sf DL} \pmb{\eta}_k\right)^{\sf H} \nonumber \\
    &~~~~ +\mathbf{I} - \mathbf{E}_k\bigg) \mathbf{\Sigma}_k \left(\mathbf{A}_k^{\sf DL}\right)^{\sf H}\bigg\Vert_F^2.
\end{align}
Then, for the convenience of notation, we define
\begin{align}
    \mathbf{M}_k &= \mathbf{\Sigma}_k\bigg(e^{j\angle\mathbf{g}_{k}^{\sf DL}}\left(e^{j\angle\mathbf{g}_{k}^{\sf DL}}\right)^{\sf H}  \nonumber \\
    &~~~~- \mathbf{Q}_k^{\sf DL} \pmb{\eta}_k\left(\mathbf{Q}_k^{\sf DL} \pmb{\eta}_k\right)^{\sf H}+\mathbf{I} - \mathbf{E}_k\bigg)\mathbf{\Sigma}_k.
\end{align}
Then, 
\begin{align}
    &\left\Vert \mathbf{A}_k^{\sf DL} \mathbf{M}_k   \left(\mathbf{A}_k^{\sf DL}\right)^{\sf H}\right\Vert_F^2 \nonumber \\
    &~~~~ = {\sf Tr}\left( \mathbf{A}_k^{\sf DL} \mathbf{M}_k   \left(\mathbf{A}_k^{\sf DL}\right)^{\sf H}\mathbf{A}_k^{\sf DL} \mathbf{M}_k^{\sf H}   \left(\mathbf{A}_k^{\sf DL}\right)^{\sf H}\right) \nonumber \\ 
    &~~~~ ={\sf Tr}\left( \left(\mathbf{A}_k^{\sf DL}\right)^{\sf H}\mathbf{A}_k^{\sf DL} \mathbf{M}_k   \left(\mathbf{A}_k^{\sf DL}\right)^{\sf H}\mathbf{A}_k^{\sf DL} \mathbf{M}_k^{\sf H}   \right) . \label{eq: Proof F1}
\end{align}
AS $N$ approaches infinity, $\lim_{N\rightarrow \infty} \frac{1}{N} \left(\mathbf{A}_k^{\sf DL}\right)^{\sf H}\mathbf{A}_k^{\sf DL} = \mathbf{I}$, we can calculate the \eqref{eq: Proof F1} as
\begin{align}
    &\lim_{N\rightarrow \infty} \frac{1}{N^2} {\sf Tr} \left( \left(\mathbf{A}_k^{\sf DL}\right)^{\sf H}\mathbf{A}_k^{\sf DL} \mathbf{M}_k   \left(\mathbf{A}_k^{\sf DL}\right)^{\sf H}\mathbf{A}_k^{\sf DL} \mathbf{M}_k^{\sf H}   \right) \nonumber \\
    &~~~~= {\sf Tr} \left( \mathbf{M}_k \mathbf{M}_k^{\sf H} \right) \nonumber \\
    &~~~~= \sum_{\ell \ne \ell'} \bigg(2\left(1 + |\eta_{k,\ell}\eta_{k,\ell'}|^2\right)|\beta_{k,\ell}\beta_{k,\ell'}|^2 \nonumber \\ 
    &~~~~~~~~-4|\eta_{k,\ell}\eta_{k,\ell'}||\beta_{k,\ell}\beta_{k,\ell'}|^2 {\rm Re} \left( e^{j \delta_{k,\ell}}e^{-j \delta_{k,\ell'}}\right)\bigg).
\end{align}
This completes the proof.

\bibliographystyle{IEEEtran}
\bibliography{abrv, reference}
\end{document}